\documentclass{emulateapj}

\usepackage{graphicx,rotate}
\newcommand{\msunyr}{\ensuremath{\mathit{M}_{\odot}{\rm yr}^{-1}}}   % msun/yr
\newcommand{\kms}{\ensuremath{{\rm km\,s^{-1}}}}                   % $\kms$ec
\newcommand{\msun}{\ensuremath{\mathit{M}_{\odot}}}   % msun
\newcommand{\lsun}{\ensuremath{\mathit{L}_{\odot}}}                  % solar luminosity
\newcommand{\rsun}{\ensuremath{\mathit{R}_{\odot}}}                  % solar radius
\newcommand{\logg}{\ensuremath{\log \mathrm{g}}}                     % log surface gravity
\newcommand{\lstar}{\ensuremath{\mathit{L}_{\star}}}                 % stellar luminosity
\newcommand{\mdot}{\ensuremath{\dot{M}}}                             % mass loss rate
\newcommand{\mstar}{\ensuremath{\mathit{M}_{\star}}}                 % stellar mass
\newcommand{\rstar}{\ensuremath{\mathit{R}_{\star}}}                 % stellar radius
\newcommand{\teff}{\ensuremath{\mathit{T}_{\rm eff}}}                % effectieve temperatuur
\newcommand{\reff}{\ensuremath{\mathit{R}_{\rm phot}}}                % effectieve temperatuur
\newcommand{\vinf}{\ensuremath{v_{\infty}}}                          % maximale uistroomsnelheid
\newcommand{\tstar}{\ensuremath{\mathit{T}_{\star}}}                 % stellar T

\newcommand{\ang}{\ensuremath{\mathrm{{\AA}}}}                % effective surface gravity
\slugcomment{ApJ, in press}

\shorttitle{On the nature of AG~Car I. Minimum phases}
\shortauthors{Groh et al.}

\begin{document}


\title{On the nature of the prototype LBV AG~Carinae\altaffilmark{1} \\ I. Fundamental parameters during visual minimum phases and changes in the bolometric luminosity during the S-Dor cycle}


\author{J. H. Groh\altaffilmark{2,3,4}}
\email{jgroh@mpifr-bonn.mpg.de}
\author{D. J. Hillier\altaffilmark{4}}
\author{A. Damineli\altaffilmark{2}}
\author{P. A. Whitelock \altaffilmark{5,6}}
\author{F. Marang\altaffilmark{5}}
\author{C. Rossi\altaffilmark{7}}

\altaffiltext{1}{Based on observations made with the 1.6m telescope at the
Observat\'orio Pico dos Dias (OPD/LNA, Brazil), with the 1.52m telescope and 8m Very Large Telescope at the European Southern Observatory (ESO, Chile), with
the International Ultraviolet Explorer (IUE) satellite, at the South African Astronomical Observatory (SAAO), and with the NASA--CNES--CSA Far
Ultraviolet Spectroscopic Explorer (FUSE), which is operated for NASA by the Johns Hopkins University under NASA contract NAS5--32985.}
\altaffiltext{2}{Instituto de Astronomia, Geof\'{\i}sica e Ci\^encias 
Atmosf\'ericas, Universidade de S\~ao Paulo, Rua do Mat\~ao 1226, Cidade 
Universit\'aria, 05508-900, S\~ao Paulo, SP, Brazil}
\altaffiltext{3}{Max-Planck-Institut f\"ur Radioastronomie, Auf dem H\"ugel 69, D-53121 Bonn, Germany (present address)}
\altaffiltext{4}{Department of Physics and Astronomy, University of Pittsburgh,
3941 O'Hara Street, Pittsburgh, PA, 15260, USA}
\altaffiltext{5}{ South African Astronomical Observatory, PO Box 9, 7935 Observatory, South Africa}
\altaffiltext{6}{ National Astrophysics and Space Science Programme, Department of Mathematics and Applied Mathematics and Department of Astronomy,
 University of Cape Town, 7701 Rondebosch, South Africa}
\altaffiltext{7}{Universit\`a di Roma "La Sapienza'', Piazzale A. Moro 5, 00185 Roma, Italy}

\begin{abstract}

We present a detailed spectroscopic analysis of the luminous blue variable AG~Carinae during the
last two visual minimum phases of its S-Dor cycle (1985--1990 and 2000--2003). The analysis reveals an
overabundance of He, N, and Na, and a depletion of H, C, and O, on the surface of AG~Car,
indicating the presence of CNO-processed material. Furthermore, the ratio N/O is higher on the
stellar surface than in the nebula. We found that the minimum phases of AG~Car are not equal to each other, since we derived a noticeable difference between the maximum effective
temperature achieved during 1985--1990 ($22,800$~K) and 2000--2001
($17,000$~K). Significant differences between the wind parameters in these two
epochs were also noticed. While the wind terminal velocity was $300~\kms$ in 1985--1990, it was
as low as $105~\kms$ in 2001.  The mass-loss rate, however, was lower from 1985--1990
($1.5\times10^{-5}~\msunyr$) than from 2000--2001 ($3.7\times10^{-5}~\msunyr$). We found that the wind
of AG~Car is significantly clumped ($f \simeq 0.10 - 0.25$) and that clumps must be formed deep
in the wind. We derived a bolometric luminosity of $1.5\times10^{6}~\lsun$ during both
minimum phases which, contrary to the common
assumption, decreases to $1.0\times10^{6}~\lsun$ as the star moves towards maximum flux in the $V$ band. Assuming that the decrease in the bolometric luminosity of AG~Car is due to the energy used to expand the outer
layers of the star (Lamers 1995), we found that the expanding layers contain roughly $0.6-2~\msun$. Such an amount
of mass is an order of magnitude lower than the nebular mass around AG~Car, but is comparable to the nebular mass found around lower-luminosity LBVs and to that of the Little Homunculus of Eta Car. If such a large amount of mass is indeed involved in the S Dor-type variability, we speculate that such instability could be a failed Giant Eruption, with several solar masses never becoming unbound from the star.   
\end{abstract}

\keywords{stars: atmospheres --- stars: mass loss --- stars: variables: other --- supergiants --- stars: individual (AG~Carinae) --- stars: rotation}
 
\section{Introduction}  \label{intro}

Luminous Blue Variables (LBVs) are characterized as high-luminosity, blue (i.e. hot) massive stars showing photometric and spectroscopic variability \citep{conti84,hd94,vg01,clark05}. AG~Carinae (=HD 94910, $\alpha_{2000}$=10:56:11.6, $\delta_{2000}$=-60:27:12.8), one of the brightest, most famous, and most variable LBVs, is considered one of the prototypes of this class, presenting all of the above characteristics.

Intense observing campaigns in the past decades have determined many of the characteristics of AG~Car. The star has photometric, spectroscopic, and polarimetric variability on a range of timescales spanning days to decades. The optical photometric variability of AG~Car was analyzed in detail by \citet{vg88}, \citet{vg90}, \citet{sterken96ibvs}, and \citet{vg97}.  These authors determined that AG~Car has strong photometric changes of up to 2.5 mag in the $V$ band on a timescale of about 5--10 years, which are attributed to the S Dor-type variability \citep{vg82,vg97b,vg01}, and {\it should not} be confused with giant eruptions similar to that experienced by Eta Carinae in the 1840's. Superimposed on these variations, $0.1-0.5~\mathrm{mag}$ changes are present on a timescale of 371 days \citep{vg97}, and photometric microvariability ($0.01-0.02~\mathrm{mag}$) with a timescale of 10--14 days has been recorded during visual minimum \citep{vg88,vg90}. \citet{whitelock83} reported the near-infrared $JHKL$ lightcurve of AG~Car between 1974--1982, showing that the amplitude of the light variations in the near-infrared is comparable to what is seen in the $V$ band. 

AG~Car is also a spectroscopic variable \citep{caputo70,humphreys70,viotti71,wolf82}, and the variability on timescale of years has been analyzed in different spectral regions: ultraviolet \citep{leitherer94,shore96}, optical \citep{leitherer94,stahl01}, and near-infrared \citep{gdj07}. During the epochs of minimum in the visual lightcurve (hereafter referred to as ``minimum"; $m_\mathrm{V}\simeq 8.1$; \citealt{vg01}), the star is relatively hot and has a WN11 spectral type \citep{sc94}, showing strong \ion{He}{1}, \ion{H}{1}, and \ion{N}{2} lines in emission, weak \ion{He}{2} 4686 $\ang$ emission, and \ion{Si}{4} 4088--4116 $\ang$  absorption \citep{stahl86,sc94,wf2000,ghd06}. During the maximum epochs of the lightcurve ($m_\mathrm{V}\sim 6.0$; \citealt{vg01}) the star is cooler, and the spectrum is reminiscent of that of extreme A-type hypergiants, with strong emission of \ion{H}{1}, \ion{Fe}{2}, and Ti II lines \citep{wolf82,stahl01}. The transition between both phases is characterized by the appearance of peculiar features in the spectrum, such as absorption-line splitting, strong electron-scattering wings in \ion{He}{1} and \ion{Fe}{2} lines, and apparent inverse P-Cygni profiles in \ion{He}{1} lines \citep{leitherer94,stahl01}. The
variable and high polarization degree measured in AG~Car motivated spectropolarimetric monitoring \citep{sl94,leitherer94,sl97,davies05,davies06a,davies08}.

The presence of a massive, bipolar circumstellar nebula around AG~Car is a testament that a recent ($t <10^4$ \,years) phase of high mass loss has occurred. The morphology and kinematics of the nebula were analyzed by \citet{nota92} and \citet{nota95}, who determined a dynamical age
of $8.5 \times 10^3$ years and a high mass of ionized nebular material ($\sim4.2~\msun$), which is likely composed of ejecta from the central star \citep{lamers01}. The nebular abundances were obtained by \citet{md90}, \citet{pacheco92}, and \citet{smith97}, showing evidence of moderate nitrogen enrichment. Properties of the circumstellar nebula were also studied in the mid- and far-infrared, revealing an incredibly high dust mass of $\sim0.25~\msun$ \citep{voors00}, dust temperature between 76--99~K \citep{voors00}, and the presence of large grains of $\sim 1$$\mu$m in order to explain the far-IR excess \citep{voors00,hyland91}. Assuming a normal gas-to-dust ratio of 100, the total nebular mass of AG~Car could be as high as $\sim 30~\msun$ \citep{voors00}, which would be of the order of, {\it or even higher than}, the mass of the Homunculus nebula around Eta Car obtained by \citet{smith03b}. At the time when the AG~Car nebula was ejected, the interstellar bubble around the central star likely contained a negligible amount of material compared to the total mass of the nebula \citep{lamers01}, implying that most of the nebular mass was ejected by the central star.

While insights into the nature of AG~Car have been obtained through the quantitative works of \citet{leitherer94} and \citet{stahl01}, the
stellar and wind parameters of AG~Car at each epoch along the S Dor-type variability cycle (hereafter S Dor cycle) are not completely constrained. Furthermore, it is currently feasible to include important physical processes in the radiative transfer codes which have had a key impact on the determination of stellar parameters of O-type (e.g. \citealt{crowther02,hillier03,bouret03,bouret05,martins05,puls06}) and Wolf-Rayet stars (e.g. \citealt{hm99,hamann06,crowther07,grafener08}). The inclusion of full line blanketing due to 100\,000s of lines, wind clumping and charge-exchange reaction, and the simultaneous treatment of the wind and photosphere in the model will considerably affect the emerging model spectrum. Together with the possibility of doing a full synthesis of the spectrum and thus using more diagnostic lines, our analysis should provide a more accurate determination of the stellar and wind properties of AG~Car.

This work is the first in a series of papers resulting from a long-term multi-wavelength spectroscopic and photometric monitoring campaign to follow AG~Car during the last two decades. In a previous paper \citep{ghd06}, we presented quantitative evidence that AG~Car had a high rotational velocity during the minimum phase of 2000--2003 and, for the first time, we showed that LBVs can be fast rotators. In this work, we present, interpret, and discuss the results obtained from observations in the last two well-documented minimum phases of AG~Car; namely, during the years of 1985--1990 and 2000--2003. Although it is subjective to define the beginning and end of minimum phases for stars which have irregular variations, such as AG~Car, we analyze in this paper epochs when the  star was fainter than $V=7$. This criterion corresponds roughly to epochs when \ion{He}{1} lines were relatively strong and displayed a classical P-Cygni profile, and time-dependent effects were minimized.

This paper is organized as follows. In Sect. \ref{agcobs1} we present the photometric and spectroscopic observations obtained during the
last two minimum phases of AG~Car, while in Sect. \ref{evolhot} we describe the spectroscopic evolution during those minima. In
Sect. \ref{agcmodel} we introduce the main characteristics of CMFGEN \citep{hm98}, the radiative transfer code used to derive the fundamental parameters of AG~Car. Sect. \ref{dist} briefly discusses the distance of AG~Car, and how the parameters are affected by that quantity. The surface chemical abundances, mass-loss rate, wind terminal velocity, stellar temperature, stellar radius, and stellar luminosity for each minimum are presented in Sect. \ref{results}, and the interpretation and discussion can be found in Sect.
\ref{discussion}. The conclusions of this paper, which suggest key changes in the current paradigm for interpreting the LBV phase and their S-Dor cycles, in particular the minimum phases, are summarized in Sect. \ref{conclusions}.

The present work will be followed by two accompanying papers. In Paper II we will discuss the evolutionary status, current mass, proximity to the Eddington limit, and whether the bi-stability mechanism is present in AG~Car. The evolution of the fundamental parameters of AG~Car during the full S Dor cycle will be analyzed, in detail, in Paper III.

\section{Observations} \label{agcobs1}

\subsection{Photometry} \label{agcphot1}

The optical photometry of AG~Car was obtained from published data from the Long-term Photometry of Variables project (LTPV;
\citealt{sterken83,manfroid91,sterken93,spoon94,manfroid95,sterken95}), accessed through the online catalog of
Vizier/CDS\footnote{http://vizier.u-strasbg.fr/viz-bin/VizieR}. The
observations were acquired using $uvby$ Str\"omgren filters on the ESO 50\,cm La Silla telescope, and the typical individual errors are between
0.003--0.006 mag. The reader is referred to the aforementioned papers for further details about the observational setup, data reduction,
and calibration. Since the LTPV project was not active after 1994, CCD $V$ magnitudes from 2000--2003 were obtained from the ASAS-3 project \citep{poj02}. The ASAS-3 data were averaged in bins of 15 days, and only high quality (grade A) data were used. Although the ASAS-3 magnitudes were measured using large apertures (75\arcsec), AG~Car is by far the dominant source of light in the field. 

For comparative purposes, estimates of the visual magnitude of AG~Car from 1981--2004 were compiled from the AAVSO archive (A. A. Heiden 2007, private communication). When both AAVSO and ASAS-3 data were available at a given epoch, their 15-day averaged magnitudes were compatible within $\sim 0.1$ mag. However, the 15-day averaged AAVSO visual lightcurve does not exactly match the more precise photoelectric $V$-band lightcurve presented by \citet{vg01}, in particular during maximum. This is not surprising, since the AAVSO sample is composed by visual estimates (i.e by eye) from different observers, and the uncertainty could be large during maximum. We estimate that the 15-day averaged AAVSO magnitudes have an uncertainty of 0.1--0.2 mag during minimum, and perhaps even more during maximum. Note that the AAVSO magnitudes are presented in this paper only for the purpose of illustrating the general behavior of AG~Car during 1981--2004, and due to their relatively low accuracy they were not used to derive a $V$-band flux. 

The near-infrared photometry was obtained using the 75cm reflector telescope from the South African Astronomical Observatory (SAAO) and
$JHKL$ filters. The SAAO photometric system is described by \citet{carter90}, and the instrument used was an $MkII$  photometer
with a 36\arcsec aperture. The typical error for each individual measurement is 0.02 mag in the $J$, $H$, and $K$ bands, and
0.05 mag in the $L$ band. In principle, the large aperture used could include some contribution from the bipolar nebula around AG~Car.
However, magnitudes obtained with diaphragm apertures of 54\arcsec, 36\arcsec, 24\arcsec, and 12\arcsec~do not present significant
differences, indicating that the nebular contribution is negligible \citep{whitelock83}. 

\begin{figure}
\resizebox{\hsize}{!}{\includegraphics{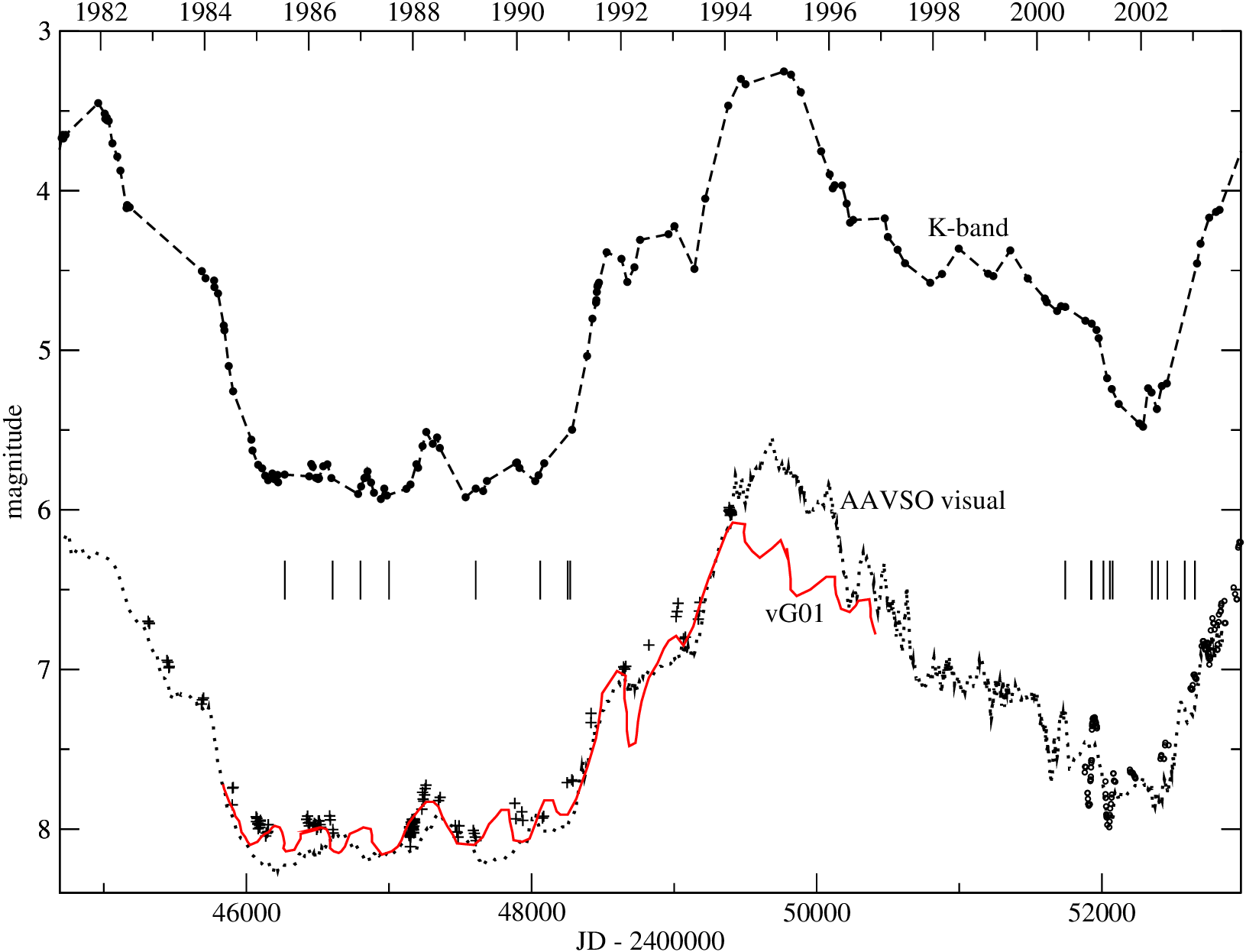}}
\caption{ AG~Car photoelectric lightcurve from 1982 to 2004 obtained in the K-band (filled circles connected by a dashed line), in the Str\"omgren $y$-band by the LTPV project (crosses; \citealt{sterken83,manfroid91,sterken93,spoon94,manfroid95,sterken95}), and in the V-band by the ASAS-3 project (open circles; \citealt{poj02}). For illustrative purposes, the schematic V-band photoelectric lightcurve presented by \citet{vg01} (red solid line) and the 15-day averaged visual magnitudes measured by the AAVSO observers (dotted line; A. A. Heiden 2007, private communication) are also shown. The vertical lines mark epochs when spectroscopic data, during minimum, were available (see Table ~\ref{obsspec1}).} \label{agcvisual}
\end{figure}

Figure \ref{agcvisual} displays the AG~Car lightcurve encompassing its last two minimum phases measured in the K-band, the photoelectric Str\"omgren $y$-band photometry obtained by the LTPV project, the ASAS-3 V-band observations, and the photoelectric lightcurve presented by \citet{vg01}. For comparison, 15-day averaged visual estimates made by the AAVSO observers are also shown. Table \ref{obsirphot1} summarizes the photometry of AG~Car used in this paper. The epochs from the first column correspond to those when spectroscopic data, during minimum, were available. The magnitudes were obtained through linear interpolation between the closest dates which had available photometry.

Fortunately, the time samplings of both the near-IR lightcurve and the optical photometry are suitable to follow the variability of AG Car on timescales of the order of months, and match the dates when the spectra were taken relatively well (Table \ref{obsirphot1}). Variability on timescales shorter than a couple of months is definitely present in AG Car \citep{vg01}, however with an amplitude of $\lesssim0.2$ mag \citep{vg88}, and thus do not affect the results obtained in this paper.

\begin{deluxetable*}{lcccccccccccccr}
%\rotate
\tabletypesize{\scriptsize}
\tablecaption{Photometry of AG Car interpolated to epochs when spectra were available during minimum\tablenotemark{a} \label{obsirphot1}}
\tablewidth{0pt}
\tablehead{ \colhead{Epoch} & \colhead{$u$}  & \colhead{$v$}  & \colhead{$b$}  & \colhead{$y$} & \colhead{visual}  & \colhead{$V$} & \colhead{$J$} & \colhead{$H$} & \colhead{$K$} & \colhead{$L$} & \colhead{$\Delta$LTPV}\tablenotemark{b}  & \colhead{$\Delta$V}\tablenotemark{b}  & \colhead{$\Delta$NIR}\tablenotemark{b} \\
&\colhead{($  0.01$)}  & \colhead{($  0.01$)} & \colhead{($  0.01$)}& \colhead{($  0.01$)} & \colhead{($  0.1$)}& \colhead{($  0.05$)}&
\colhead{($  0.02$)} &\colhead{($  0.02$)}&\colhead{($  0.02$)}&\colhead{($  0.05$)}& \colhead{(days)}& \colhead{(days)}& \colhead{(days)}
}
\startdata
%    Date              u        v        b        y     AAVSO     ASAS V     J         H        K        L     dLTPV     dASAS   dIR 
1985  Jul	19   &  8.94  &  8.72  &  8.37  &  7.96  &  8.2   & \ldots &  6.45  &  6.13  &  5.78  &  5.34  &  112   & \ldots &  -3  \\ 
1986  Jun	17    &  9.04  &  8.78  &  8.41  &  7.98  &  8.1   & \ldots &  6.47  &  6.12  &  5.80  &  5.35  &    $-7$   & \ldots &  1   \\
1986  Jun	18    &  9.05  &  8.79  &  8.41  &  7.98  &  8.1   & \ldots &  6.47  &  6.12  &  5.80  &  5.35  &    $-6$   & \ldots &  2   \\
1986  Jun	23    &  9.07  &  8.81  &  8.43  &  8.00  &  8.1   & \ldots &  6.47  &  6.13  &  5.81  &  5.36  &    $-1$   & \ldots &  7   \\
1987  Jan	05    &  9.03  &  8.81  &  8.43  &  8.03  &  8.1   & \ldots &  6.52  &  6.19  &  5.86  &  5.36  &  192   & \ldots &  $-4$   \\
1987  Jun	10    &  9.00  &  8.81  &  8.43  &  8.03  &  8.2   & \ldots &  6.57  &  6.24  &  5.93  &  5.63  &  $-183$   & \ldots & 12  \\  
1987  Jul	24    &  8.99  &  8.80  &  8.43  &  8.03  &  8.1   & \ldots &  6.57  &  6.23  &  5.91  &  5.50  &  $-139$   & \ldots & 14  \\  
1989  Mar	26    &  8.98  &  8.83  &  8.45  &  8.07  &  8.2   & \ldots &  6.54  &  6.19  &  5.87  &  5.22  &    1   & \ldots &  2  \\  
1989  Dec   23    &  8.79  &  8.63  &  8.26  &  7.85  &  8.1   & \ldots &  6.38  &  6.05  &  5.71  &  5.25  &    1   & \ldots & $ -8 $ \\  
1990  Apr	30    &  8.84  &  8.72  &  8.36  &  7.94  &  7.9   & \ldots &  6.42  &  6.11  &  5.81  &  5.37  &   $-64$   & \ldots & $-15$  \\  
1990  Jun	16    &  8.88  &  8.72  &  8.35  &  7.93  &  8.0   & \ldots &  6.41  &  6.08  &  5.77  &  5.32  &   $-17$   & \ldots &  8  \\
1990  Jun	18    &  8.88  &  8.72  &  8.35  &  7.93  &  8.0   & \ldots &  6.41  &  6.07  &  5.76  &  5.32  &   $-15$   & \ldots & 10   \\ 
1990  Aug	07    &  8.83  &  8.67  &  8.29  &  7.89  &  8.0   & \ldots &  6.32  &  5.99  &  5.69  &  5.24  &   26   & \ldots & 20  \\ 
1990  Dec   22    &  8.66  &  8.51  &  8.14  &  7.71  &  8.0   & \ldots &  6.18  &  5.86  &  5.54  &  5.12  &    $-1$   & \ldots &  $-37$   \\ 
1990  Dec   28    &  8.66  &  8.50  &  8.13  &  7.71  &  8.0   & \ldots &  6.17  &  5.85  &  5.53  &  5.12  &    4   & \ldots & $-31 $ \\
1991  Jan	21    &  8.65  &  8.48  &  8.12  &  7.69  &  7.9   & \ldots &  6.15  &  5.83  &  5.51  &  5.10  &    $-4$   & \ldots & $ -7 $ \\
2000  Jul	18    & \ldots & \ldots & \ldots & \ldots &  7.3   & \ldots &  5.54  &  5.15  &  4.73  &  4.18  & \ldots & \ldots &  0  \\
2000  Dec   12        & \ldots & \ldots & \ldots & \ldots &  7.5   & 7.67   &  5.58  &  5.20  &  4.82  &  4.29  & \ldots &     2  &  5  \\
2001  Jan	17    & \ldots & \ldots & \ldots & \ldots &  7.5   & 7.57   &  5.57  &  5.21  &  4.83  &  4.32  & \ldots &     0  & $ -3 $ \\
2001  Apr	12    & \ldots & \ldots & \ldots & \ldots &  7.7   & 7.63   &  5.80  &  5.45  &  5.07  &  4.55  & \ldots &    $-13$  & $ -26 $ \\
2001  May	27    & \ldots & \ldots & \ldots & \ldots &  7.9   & 7.90   &  5.91  &  5.58  &  5.21  &  4.71  & \ldots &     0  &$ -14$  \\
2001  Jun	10    & \ldots & \ldots & \ldots & \ldots &  7.9   & 7.78   &  5.94  &  5.61  &  5.24  &  4.75  & \ldots &     0  &  0  \\
2001  Jun	15    & \ldots & \ldots & \ldots & \ldots &  7.9   & 7.71   &  5.95  &  5.62  &  5.25  &  4.76  & \ldots &    $ -4 $ &  4  \\
2002  Mar	17    & \ldots & \ldots & \ldots & \ldots &  7.7   & 7.60   &  5.93  &  5.60  &  5.27  &  4.93  & \ldots &  $ -60 $ &  1  \\
2002  Apr	30    & \ldots & \ldots & \ldots & \ldots &  7.8   & 7.57   &  6.01  &  5.68  &  5.34  &  4.92  & \ldots &  $  -16 $ &  7  \\
2002  Jul	04    & \ldots & \ldots & \ldots & \ldots &  7.8   & 7.47   &  5.88  &  5.54  &  5.20  &  4.79  & \ldots &   $  -8 $ &  2  \\
2002  Jul	20    & \ldots & \ldots & \ldots & \ldots &  7.7   & 7.46   &  5.84  &  5.49  &  5.14  &  4.73  & \ldots &     7  & 18  \\
2002  Nov   04        & \ldots & \ldots & \ldots & \ldots &  7.2   & 7.20   &  5.51  &  5.12  &  4.76  &  4.30  & \ldots & $ -39$  & $ -85$  \\
2003  Jan	11    & \ldots & \ldots & \ldots & \ldots &  7.2   & 7.03   &  5.31  &  4.89  &  4.52  &  4.03  & \ldots &     0  & $ -17$  \\
\enddata
\tablenotetext{a}{Photometric errors are indicated in parenthesis. References: $uvby$ Str\"omgren=LTPV project \citep{sterken83,manfroid91,sterken93,spoon94,manfroid95,sterken95}; AAVSO visual=A. A. Heiden 2007, private communication; ASAS V-band=\citet{poj02}; JHKL=SAAO (\citealt{whitelock83}; this work).}
\tablenotetext{b}{Time delay between the epochs when spectroscopic and photometric observations were obtained (JD$_\mathrm{spec} -  $JD$_\mathrm{phot}$).}
\end{deluxetable*} 

\subsection{Spectroscopy} \label{agcspec1}

\subsubsection{Minimum of 1985--1990}

Flux-calibrated ultraviolet spectra of AG~Car were obtained by the {\em International Ultraviolet Explorer} (IUE) satellite in the region 1150--3120 $\ang$ and gathered through the {\em Space Telescope Science Institute} (STSCI) public-data archive.
The IUE absolute-flux calibration is precise to about 5\% (degrading in the edges of the spectral range), which is adequate for our
purposes. AG~Car was observed in various epochs during the 1985--1990 minimum with the SWP and LWP high-resolution cameras, with
$ R \simeq 15,000$ in the wavelength range of 1150--3200 $\ang$ (Table \ref{obsspec1}, see also \citealt{leitherer94,shore96}).

Optical spectra of AG~Car were obtained using different telescopes in the southern hemisphere. The Echelle spectrograph
mounted on the 1.4m CAT/ESO (Chile) telescope was used to collect high-resolution spectra from 1986 June to 1991 January around
H$\alpha$, \ion{He}{1} 5876 $\ang$, and \ion{He}{2} 4686 $\ang$, with R=60,000. Most of these data have been described elsewhere \citep{bandiera89,viotti91,leitherer94}. Lower resolution data between 3800--4900 $\ang$ were obtained using the 1m CTIO/Yale telescope and were discussed by \citet{wf2000}. Additional lower-resolution observations using the longslit Cassegrain spectrograph on the 1.52m ESO/La Silla telescope were made on 1990 June 18--19 between 4330--4870 $\ang$ (R=2300) and 6200--7270 $\ang$ (R=1700), and were described by \citet{viotti93}. 

It was not possible to obtain a precise continuum normalization of the Echelle data around H$\alpha$ since the spectral coverage was very small and the electron scattering wings were very extended. Therefore, we scaled the Echelle spectra in flux in order to match the electron-scattering wings of the longslit data, as described by \citet{leitherer94}. We conservatively estimate that the uncertainty in the continuum after this procedure is at most $\sim 5\%$. The same procedure could not be applied to the Echelle spectra around \ion{He}{1} 5876 $\ang$ and, as a consequence, the strength of the electron scattering wings are not reliable, and are likely underestimated, around this line. 

Additionally, the near-infrared region around 9950--10200 $\ang$ was covered by CCD observations made on the 1.6m telescope of the
Observat\'orio Pico dos Dias (OPD/LNA, Brazil) using the Coude spectrograph. The spectra typically have $R \simeq 10,000$, and the data reduction
was accomplished using standard IRAF tasks for bias subtraction, flat-field correction, extraction of the spectrum, wavelength
calibration, and continuum normalization. Those spectra were also published by \citet{leitherer94}, where further details can be found.

\subsubsection{Minimum of 2000--2003}

During this minimum we used the Fiber-fed Extended Range Optical
Spectrograph (FEROS; \citealt{kaufer99}) mounted on the 1.52 m telescope of ESO/La Silla (Chile) to obtain spectra between 2001 January
and 2002 March. The instrumental configuration provided spectral coverage between 3600--9200 $\ang$ at R=48,000, and the data was reduced
using the FEROS pipeline \citep{stahl99}. We also used data obtained on 2003 January 11 with the UV-visual
Echelle Spectrograph (UVES) mounted on the ESO 8m Very Large Telescope (VLT, program 266.D-5655, \citealt{bagnulo03}). The resolution
provided by UVES was $R=80,000$ in the wavelength range 3000--10400 $\ang$.

We used the near-infrared camera CamIV on the 1.6m telescope of OPD/LNA (Brazil) to obtain long-slit spectra of AG~Car in the
wavelength range 10450--11000 $\ang$ ($R=7000$) in 2001 June, 2002 July, and 2002 November. Further details on the observational setup and
data reduction are given in \citet{gdj07}, where the data are also discussed. 

With the decommissioning of IUE, no ultraviolet monitoring of the star was possible during the 2000--2003 minimum. Fortunately,
a {\it Far Ultraviolet Spectroscopic Explorer} (FUSE) visit occurred in 2001 May 27, covering the far-ultraviolet region between 900--1180 $\ang$ ($R=20,000$). The FUSE spectrum was obtained from the online FUSE public-data archive.

The spectroscopic observations obtained during minimum are summarized in Table \ref{obsspec1}.

%TABLE 2
 
\begin{deluxetable}{lccr}
\tabletypesize{\scriptsize}
\tablecaption{Journal of spectroscopic observations of AG~Car during minimum \label{obsspec1}}
\tablewidth{0pt}
\tablehead{ \colhead{Date} & \colhead{Telescope} & \colhead{Spectral Range ($\ang$)} & \colhead{$R$} }
\startdata
1985 July 19 & IUE SWP & 1180--1900 & 18,000 \\
1986 June 17 & 1.4m ESO & 5860-5910 & 60,000\\
1986 June 18 & 1.4m ESO & 6532-6588 & 60,000\\
1986 June 23 & IUE SWP & 1180--1900 & 18,000\\
1986 June 23 & IUE LWP & 1800--3150 & 13,000 \\
1987 January 05 & 1.4m ESO & 5860-5910 & 60,000\\
1987 January 05 & 1.4m ESO & 6532-6588 & 60,000\\
1987 June 10 & 1.4m ESO &  3917-3947 & 60,000\\
1987 June 10 & 1.4m ESO & 6532-6588 & 60,000\\
1987 July 24 & IUE SWP & 1180--1900 & 18,000 \\
1987 July 24 & IUE LWP & 1800--3150 & 13,000 \\
1989 March 26 & 1m CTIO/Yale  & 3800-4900 & 3000\\
1989 December 23 & IUE SWP & 1180--1900 & 18,000 \\
1989 December 23 & IUE LWP & 1800--3150 & 13,000 \\
1990 April 30 & IUE SWP & 1180--1900  & 400 \\
1990 June 16 & 1.6m LNA & 9950--10200 & 10,000 \\
1990 June 18 & 1.52m ESO & 4330--4870 & 2300 \\
1990 June 18 & 1.52m ESO & 6200--7270 & 1700 \\
1990 August 07 & IUE LWP & 1800--3150  & 270 \\
1990 December 22 & 1.6m LNA & 9970--10080 & 10,000\\
1990 December 28 & 1.6m LNA & 6490--6710 & 10,000 \\
1991 January 21 & 1.4m ESO & 4550--4750 & 60,000 \\
2000 July 18 & 1.6m LNA & 9950--10200 & 10,000 \\
2000 July 18 & 1.6m LNA & 6350--7100 & 10,000 \\
2000 December 12 & 1.6m LNA & 6350--7100 & 10,000 \\
2001 January 17 & 1.52m ESO & 3600--9200 & 48,000 \\
2001 April 12 & 1.52m ESO & 3600--9200 & 48,000 \\
2001 May 27 & FUSE & 900-1180 & 20000 \\
2001 June 10 & 1.6m LNA & 10450--11000 & 7000 \\
2001 June 15 & 1.52m ESO & 3600--9200 & 48,000 \\
2002 March 17 & 1.52m ESO & 3600--9200 & 48,000 \\
2002 April 30 & 1.6m LNA & 10450--11000 & 7000 \\
2002 July 04 & 1.52m ESO & 3600--9200 & 48,000 \\
2002 July 20 & 1.6m LNA & 10450--11000 & 7000 \\
2002 November 04 & 1.6m LNA & 10450--11100 & 7000 \\
2003 January 11 & 8m ESO/VLT & 3000-10400 & 80,000\\
\enddata
\end{deluxetable}

\section{Spectroscopic variability of AG~Car during minimum } \label{evolhot}

In this Section we present the spectroscopic evolution of AG~Car during minimum epochs. Since other WN9--WN11 stars also display variability in the spectrum on timescales of years \citep{mcgregor88,crowther97,gdj07}, investigating the variability behavior of AG~Car might provide insights on the evolutionary link between LBVs and WN9--WN11 stars and on the physical mechanism which drives the S-Dor variability.

\subsection{Minimum of 1985--1990}  \label{evolhot90}

\citet{leitherer94} monitored the ultraviolet 
spectrum during the minimum of 1985--1990 and concluded that little or no variability was present. Indeed, the low level of variability is consistent with the visual lightcurve (Fig. \ref{agcvisual}), which shows only low-amplitude photometric variability of $\sim 0.1-0.3$ mag  during 1985--1990 \citep{vg88,vg90,leitherer94,stahl01}. The evolution of the optical spectrum from the end of the minimum (1990 December) until epochs on the rise to the lightcurve maximum (1992 June) was also followed by \citet{leitherer94}.

\begin{figure}[!t]
\resizebox{\hsize}{!}{\includegraphics{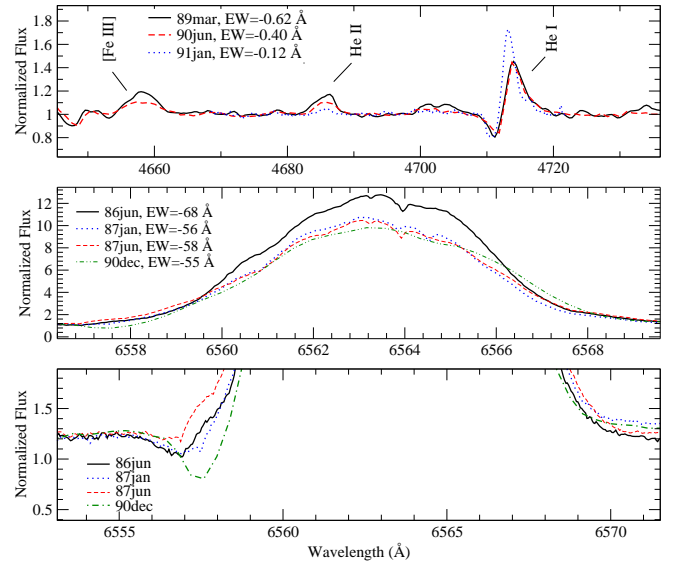}} 
\caption{\label{varhalpha90} {\it Top}: AG~Car spectrum around [\ion{Fe}{3}] 4657 $\ang$ and \ion{He}{2} 4686 $\ang$, obtained in 1989 March (black solid line), 1990 June (red dashed), and 1991 January (blue dotted). The resolution of the spectra obtained in 1989 March and 1991 January were adjusted to the same resolution as the spectrum from 1990 June ($R=2300$). Quoted EWs refer to \ion{He}{2} 4686 $\ang$, and have an error of $\sim 0.03$ $\ang$. {\it Middle}: AG~Car spectrum around H$\alpha$ obtained in 1986 June (black solid line),  1987 January (blue dotted), 1987 June (red dashed), and 1990 December (green dot-dashed). Quoted EWs refer to the emission component of H$\alpha$, and have an error of $\sim 2$ $\ang$ which is dominated by the uncertainty in the continuum level. {\it Bottom}: Zoom in around the variable P-Cygni absorption component that appears during some
epochs, indicating changes in the ionization structure of hydrogen in the outer wind, as suggested by \citet{crowther97} to explain a similar behavior seen
in the LBV candidate He3-519. The offset of the continuum level from 1.0 is due to the strong H$\alpha$ electron scattering wings.}
\end{figure}

Here we draw our attention to the evolution of the optical spectrum and photometry inside the 1985--1990 minimum. Spectra around \ion{He}{2} 4686 $\ang$ were available only during 1989--1991 and are shown in Fig. \ref{varhalpha90}. These data clearly show that \ion{He}{2} 4686 $\ang$ became progressively weaker from 1989 until 1991, indicating a decrease of the effective temperature. The high-ionization line of [\ion{Fe}{3}] 4657 $\ang$ also became slightly weaker from 1989 March until 1990 June.  

The H$\alpha$ line was followed with a better time sampling than \ion{He}{2} 4686 $\ang$ (Fig. \ref{varhalpha90}). It can be noticed that moderate variations were present in the emission component of H$\alpha$, with a decrease in the  equivalent width (EW) from 1986 June to 1987 January. The EW of H$\alpha$ varied from $-68 \pm 2$ $\ang$ in 1986 June to $-56 \pm 2~\ang$ in 1987 January (Fig. \ref{varhalpha90}), and then remained approximately constant until 1990 December. On the other hand, the $y$-band magnitude was 8.00 mag in 1986 June, remained roughly constant during 1987, and only changed significantly to 7.71 mag in 1990 December (Fig. \ref{agcvisual}). Therefore, the changes in the emission component in H$\alpha$ can be explained by a net increase in the continuum emission compared to the line emission only in 1990 December -- on the other hand, the line emission was indeed stronger in 1986 June compared to the rest of the minimum. 

In addition to the changes in the emission component, a weak P-Cygni absorption in H$\alpha$ developed during some epochs between 1985--1990. Interestingly, a
similar behavior was previously noticed in the LBV candidate He 3-519 \citep{crowther97}, whose spectrum is remarkably similar to AG~Car's during the 1985--1990
minimum \citep{sc94,wf2000}. \citet{crowther97} suggested that this behavior is due to changes in the ionization structure of hydrogen in the outer parts of the wind. In this scenario, the absorption component seen in H$\alpha$ appears when a fraction of hydrogen recombines in the outer part of the wind and disappears when hydrogen is mainly ionized. A similar behavior has also been detected in the LBV P Cygni \citep{pauldrach90,najarro97}. 

Comparing the evolution of the emission and absorption components of H$\alpha$, we suggest that the stellar and/or wind parameters are variable even during minimum epochs, and radiative transfer models are needed to quantify these changes. The detailed analysis of the 1985--1990 minimum spectra will be presented in Sect. \ref{jun90}. 

\subsection{Minimum of 2000--2003}

\begin{figure*}
\resizebox{\hsize}{!}{\includegraphics{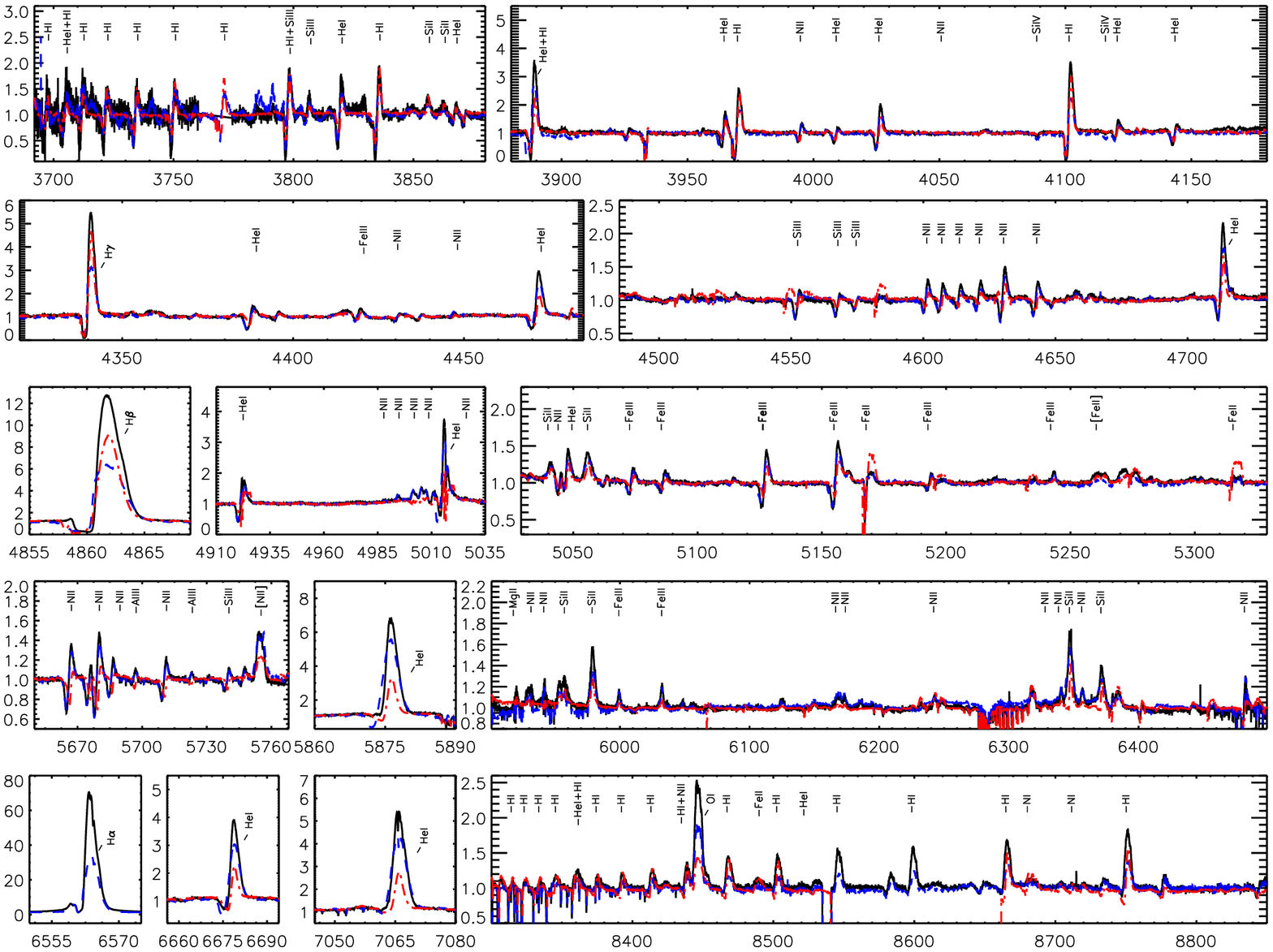}} 
\caption{\label{evolopt0103a} Evolution of the optical spectrum of AG~Car in the region $3690-8850~\ang$ during the 2000--2003 minimum (see text for discussion). The spectra were obtained in 2001 April (black solid line), 2002 March (blue dashed line), and 2003 January (red dot-dashed line). Data obtained from 2000 July until 2001 June show no significant variability. The spectrum from 2003 January has a small gap in wavelength coverage between $8540-8660~\ang$, while H$\alpha$ is saturated and not shown.}
\end{figure*}

Figure \ref{evolopt0103a} displays the evolution of the optical spectrum of AG Car between 3690--8850 $\ang$. The variability in the near-infrared region from $10500-11000~\ang$ was presented by \citet{gdj07}.

In general, the spectral lines became broader from 2001 April to 2002 March and then narrower in 2003 January, indicating changes in the wind terminal velocity as a
function of temperature. That behavior is interpreted as due to the fast rotation of AG~Car \citep{ghd06}, but might also be caused by the time-dependent
nature of the velocity field. The hydrogen Balmer and Paschen lines (Fig. \ref{evolopt0103a}) show a significant change in strength and line profile inside the minimum: they became weaker from 2001 April to 2002 March and then stronger in 2003 January, indicating changes in the wind density. \ion{He}{2} 4686 $\ang$ is very weak during the minimum of 2000--2003, with EW$\simeq-100 \pm10~\mathrm{m{\AA}}$ in 2001 April and 2001 June spectra, and EW$\simeq-80 \pm30~\mathrm{m{\AA}}$ in the 2002 March spectrum. \ion{He}{2} 4686 $\ang$ is absent in the 2003 UVES spectrum, with an upper limit of $15~\mathrm{m{\AA}}$. Comparatively, this line is readily identified in the spectrum of AG~Car during the 1985--1990 minimum (Fig. \ref{varhalpha90}; 
\citealt{stahl86,viotti93,leitherer94,sc94,wf2000,stahl01}). 

Meanwhile, the \ion{He}{1}, \ion{Fe}{3}, and \ion{N}{2} lines became weaker from 2001 to 2003, while the very weak \ion{Fe}{2} and \ion{Mg}{2} lines seen in the 2001 and 2002 spectra are prominent in the data from 2003. This behavior indicates that the effective temperature was decreasing from 2001 to 2003. Therefore, the above picture is consistent with the star moving out of minimum from 2001 to 2003, as suggested by the visual lightcurve (Fig. \ref{agcvisual}).

\begin{figure}
\resizebox{0.95\columnwidth}{!}{\includegraphics{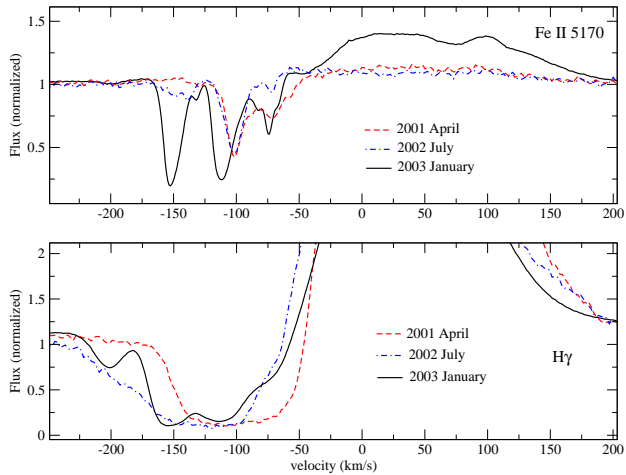}} 
\caption{\label{agchotabs}Evolution of the multiple absorption components seen in the AG~Car spectrum during the minimum of 2000--2003.
To illustrate their general behavior, we display the \ion{Fe}{2} 5170 $\ang$ line (top) and H$\gamma$ 4340 $\ang$ (bottom). The high-velocity multiple absorption components are likely caused by time-dependent effects, while the $-72~\kms$ component may also be formed in the circumstellar nebula of AG~Car, which has an expansion velocity of $\sim 70~\kms$ \citep{nota92}.}
\end{figure}

Interestingly, multiple absorption components appeared during 2001--2003 in the hydrogen, \ion{Fe}{2}, and \ion{Ni}{2} lines (Fig. \ref{agchotabs}), and are reminiscent of what was seen during 1991--1993 \citep{leitherer94,stahl01} when the star was also moving towards maximum. Three absorption components in the \ion{Fe}{2} lines can be securely identified in the data from 2000--2003, and they are centered at $-72$, $-101$, and $-148~\kms$ in the FEROS spectra from 2001--2002. These absorption components change in velocity and in strength in timescales as short as 1 year, as first noted by \citet{stahl01}. Our data also present a similar behavior, and in the 2003 UVES spectrum, these absorption components shift to $-74$, $-112$, and $-153~\kms$. An additional absorption component centered at $-202~\kms$ is also present in the \ion{Fe}{2} and hydrogen Balmer lines in 2003. 

The $-72~\kms$ component could arguably be formed in the distant circumstellar nebula, which has an expansion velocity of $\sim 70~\kms$ \citep{nota92}, but that component is not observed at all epochs (\citealt{leitherer94,stahl01}, this work). The other multiple absorption components are unlikely to be formed in the distant circumstellar nebula of AG~Car, since they change in velocity and are approximately centered at the wind terminal velocity found in previous epochs. Thus, the multiple absorption components in AG~Car are likely not reminiscent of the multiple ejecta absorption components seen in the massive LBV Eta Carinae \citep{gull05,nielsen05,johansson05,gull06}. We support the scenario, proposed by \citet{stahl01}, that the multiple-absorption features are formed much closer to the star and are due to regions in the wind which have a density decrease (or enhancement) due to the temporal variability of the star. Indeed, this behavior is predicted by detailed radiative transfer models which include time-dependent effects (\citealt{gdh08}; J. H. Groh et al. 2009, in preparation).

\section{The Model} \label{agcmodel}

To analyze the spectra of AG~Car, we used the radiative transfer code CMFGEN \citep{hillier87,hillier90,hm98,hm99,bh05}, which has been successfully used to model the spectrum of LBVs and related objects \citep{sc94,najarro94,najarro97,najarro09,najarro01,hillier98,hillier01,hillier06,figer98,drissen01,bresolin02,marcolino07}. The code has been extensively discussed in the aforementioned references, and below we concisely describe the main characteristics of the code. 

CMFGEN assumes an spherical-symmetric, steady-state outflow, and computes continuum and line formation in the non-LTE regime. Each model is defined by the hydrostatic stellar
radius $\rstar$, the luminosity L$_\star$, the mass-loss rate $\dot{M}$, the wind terminal velocity $\vinf$, the stellar mass $M$, and by the abundances Z$_i$ of the included species. We define the hydrostatic radius $\rstar$  as the radius where the velocity is $v(\rstar)=v_\mathrm{sonic}/3$, in order to avoid any effects due to the strong wind in determination of $\rstar$, which is the case for $v \gtrsim v_\mathrm{sonic}/3$. Thus, the velocity of the hydrostatic core, where $\rstar$ and $\tstar$ are defined, is typically $4-5~\kms$ for AG~Car during minimum. 

CMFGEN does not solve the hydrodynamic equations of the wind, and therefore, a velocity law
$v(r)$ has to be assumed a priori. In CMFGEN, the velocity law is parameterized either by a single beta-type law or by a two-beta
velocity law which is modified at depth to reach a hydrostatic structure.  The stellar mass $M$ is needed to compute the density structure below and close to the sonic point, and therefore changes in $M$ might affect the spectroscopic analysis. In this paper we assume $M = 70~\msun$, since lower values of $M$ would imply that the star is above the Eddington limit modified by rotation (Paper II). A thorough discussion on the impact of $M$ on the spectral analysis will be provided in Paper II. The velocity structure below the sonic point is iterated to fulfill the wind momentum equation within 10\% for $v \lesssim 8~\kms$. A beta-type law is joined to this structure just below the sonic point. Although the code solves for a hydrostatic structure, there might still be uncertainties in the density structure below the sonic point due to the proximity to the Eddington limit and the choice of $M$, rotation rate, and viewing angle. 

CMFGEN allows for the presence of clumping within the wind using a volume-filling factor approach. The wind is assumed to be homogenous close to
the star, become partially clumpy at a characteristic velocity scale $v_c$, and achieves full clumpiness, with a volume-filling factor $f$, at large distances, as follows: \begin{equation}
f(r)=f+(1-f)\exp[-v(r)/v_c]\,\,.
\end{equation} 

Full line blanketing is included consistently in CMFGEN through the concept of superlevels, which groups similar energy levels
into one single superlevel to be accounted for in the statistical equilibrium equations. In the analysis of AG~Car, our initial atomic model
consisted of H, He, C, N, O, and Fe in order to constrain the parameter regime to be explored. For the final modeling we included additional elements and ionization stages. The atomic models adopted for the final modeling are summarized in Table \ref{agcmodatom}, and included H, He, C, N, O, Na, Mg, Al, Si, Cr, Mn, Fe, Co, and Ni.

The synthetic spectra of AG~Car presented in this paper were computed using {\sc CMF\_FLUX} \citep{hm98,bh05} and did not include effects due to rotation. As discussed in \citet{bh05}, rotation mainly affects spectral lines formed close to the photosphere. Specifically in the case of AG~Car during minimum, only \ion{Si}{4} and \ion{He}{2} lines have their profile significantly changed by rotation (\citealt{ghd06}; Paper II).

\begin{deluxetable}{lccr}
\tabletypesize{\footnotesize}
\tablecaption{Final atomic model used in the analysis of AG~Car \label{agcmodatom}} 
\tablewidth{0pt}
\tablehead{\colhead{Ion} &  \colhead{$N_\mathrm{S}$} & \colhead{$N_\mathrm{F}$} & \colhead{$N_\mathrm{T}$}}
\startdata
\ion{H}{1}& 20 & 30 & 435 \\
\ion{He}{1} & 40 & 45 & 233 \\ 
\ion{He}{2} & 22 & 30 & 435 \\
\ion{C}{2} & 16 & 54 & 366 \\
\ion{C}{3} & 30 & 54 & 268 \\
\ion{C}{4} & 21 & 64 & 1446 \\
\ion{N}{1} & 44 & 104 & 885 \\
\ion{N}{2} & 41 & 84 & 530 \\ 
\ion{N}{3} & 30 & 63 & 344 \\
\ion{O}{1} & 69 & 161 & 358 \\
\ion{O}{2} & 19 & 57 & 372 \\
\ion{O}{3} & 17 & 79 & 450 \\
\ion{Na}{1} & 22 & 71 & 1550 \\ 
\ion{Mg}{2} & 18 & 45 & 358 \\
\ion{Al}{2} & 21 & 65 & 1348 \\
\ion{Si}{2} & 22 & 43 & 182 \\
\ion{Si}{3} & 20 & 34 & 92 \\
\ion{Si}{4} & 22 & 33 & 183 \\
\ion{Cr}{2} & 42 & 310 & 2127 \\
\ion{Cr}{3} & 40 & 209 & 4077 \\
\ion{Cr}{4} & 29 & 234 & 4466 \\
\ion{Cr}{5} & 30 & 223	& 3207 \\
\ion{Mn}{2} & 33 & 333 & 1348 \\
\ion{Mn}{3} & 33 & 333 & 4038 \\
\ion{Mn}{4} & 39 & 464 & 11103 \\
\ion{Mn}{5} & 16 & 80 & 676  \\
\ion{Fe}{2} & 61 & 261 & 3837 \\
\ion{Fe}{3} & 33 & 191 & 6935 \\
\ion{Fe}{4} & 58 & 383 & 2816 \\
\ion{Fe}{5} & 25 & 168 & 1938 \\
\ion{Co}{2} & 54 & 421 & 3267 \\
\ion{Co}{3} & 22 & 174 & 3837 \\ 
\ion{Ni}{2} & 29 & 204 & 4728 \\
\ion{Ni}{3} & 28 & 220 & 5181 \\
\ion{Ni}{4} & 36 & 200 & 2337 \\
\ion{Ni}{5} & 46 & 183 & 1524 \\
\enddata
\tablecomments{For each species, an additional one-level ion corresponding to the highest ionization stage was included explicitly in the rate equations (i.e. \ion{C}{5} for carbon, for instance), but are not shown above for brevity. The columns correspond to  the number of included superlevels ($N_\mathrm{S}$), total energy levels ($N_\mathrm{F}$), and the number of bound-bound transitions included for each ion ($N_\mathrm{T}$). } 
\end{deluxetable}

\section{Distance and systemic velocity}  \label{dist}

The spectroscopic analysis using CMFGEN cannot constrain the distance to AG~Car. Throughout this work, we assume that AG~Car is located at $d=6~\mathrm{kpc}$ \citep{humphreys89,hoekzema92}.

\citet{stahl01} re-determined the heliocentric systemic velocity of AG~Car and found $v_{\mathrm{sys}}=10 \pm 5~\kms$, suggesting a lower kinematical distance in the range of 5--6 kpc, instead of a kinematical distance of 6.0--6.5 kpc from \citealt{humphreys89}\footnote{We updated the values of the kinematical distance presented by \citet{humphreys89}, $d_\mathrm{kin}\simeq6.4 - 6.9$ kpc, in order to account for a revised galactocentric distance of the Sun of $R_0=8$~kpc and $\Theta_0=220~\kms$, obtaining $d_\mathrm{kin}\simeq6.0-6.5$ kpc, depending on whether the Galactic rotation curve of \citet{fich89} or \citet{bb93} was chosen.}. Taking the errors into account, both determinations are consistent \citep{stahl01}. We also measured the centroid of the [\ion{Fe}{2}] and [\ion{N}{2}] lines in our high-resolution data and obtained $v_{\mathrm{sys}}$ similar to the value found by \citet{stahl01}. 

A note of caution regarding the kinematical distance of early-type stars should be pointed out: the distance to young stellar clusters based on spectroscopic parallax are systematically lower than their kinematical distance (e.g., \citealt{bcd00,bdc01,fbdc02,fbdc08}). In addition, about 10--30 \% of Galactic O-type stars are runaways \citep{gies87,dewit05}, implying that their derived kinematical distance might be meaningless. However, in the case of AG~Car, the empirical luminosity-amplitude relationship found for LBVs \citep{wolf89} and the distance versus reddening relationship of nearby stars both support a distance of 6 kpc \citep{humphreys89}. 

The bolometric luminosity $\lstar$, stellar radius $\rstar$, and mass-loss rate $\mdot$ obtained through the spectroscopic analysis are distance-dependent, while the effective temperature and chemical abundances are only weakly, if at all, dependent on $d$. Following \citet{schmutz89}, \citet{najarro97}, and \citet{hillier98}, the physical parameters of AG~Car obtained in this work (see Sect. \ref{results}) should be scaled to other distances according to: $\lstar \propto d^{2}$, $\mdot \propto d^{1.5}$, $\rstar \propto d$, and $\teff \propto d^0 \,.$

\section{Results } \label{results}

\subsection{Surface chemical abundances } \label{agcabund}

The surface chemical abundances obtained for AG~Car are listed in Table \ref{abund_table}. The values obtained are consistent
when comparing different epochs from both minima, supporting the physical parameters derived for those epochs (Sect. \ref{jun90} and \ref{2001-03}).

The helium abundance of AG~Car was determined as He/H=$0.43 \pm 0.08$ (in number) in order to reproduce the intensity of \ion{He}{1}, \ion{He}{2} and hydrogen lines in the optical and near infrared spectrum. This value is in agreement with previous works  \citep{sc94,leitherer94,stahl01}, and with the helium content for the LBV phase predicted by evolutionary models \citep{meynet00}. Unlike Eta Carinae \citep{hillier01} and HD 316285 \citep{hillier98}, the deduced helium abundance of AG~Car does not scale with the derived mass-loss rate. This is because AG~Car simultaneously presents \ion{He}{1} and \ion{He}{2} lines in the spectrum during minimum, while Eta Car and HD 316285 do not. 

In addition to the helium content, it is essential to determine the abundance of the CNO-elements to constrain the
evolutionary stage of AG~Car. A nitrogen mass fraction of $(7.2\pm2)\times10^{-3}$ ($11.5 \pm 3.4 $ times the solar value;  hereafter, we used the solar abundance 
values listed in \citealt{grevesse07} and references therein) was obtained based on numerous \ion{N}{2} lines in the optical spectrum (4601--4643, 6482 $\ang$) and on ultraviolet \ion{N}{3} lines. The carbon abundance was determined based on the strength of optical \ion{C}{2} lines at 6578--6582 $\ang$ and 7231--36 $\ang$, yielding a carbon
mass fraction of $(2.4\pm0.7)\times10^{-4}$ ($0.11 \pm 0.03$ times the solar value). The errors in the carbon and nitrogen abundance are about 30\%.
The oxygen mass fraction was determined to be $(2.4\pm1)\times10^{-4}$ ($0.04 \pm 0.02$ times the solar value) using the O I lines at 7774--8446 $\ang$
and \ion{O}{2} 4620 $\ang$. As it was not possible to fit both O I lines with the same model, the uncertainty in the oxygen abundance is
around 50\%. Nonetheless, it is safe to point out that the oxygen abundance on the surface is very low compared to the solar value. 
Together with the high nitrogen content and carbon depletion, we conclude that there is certainly CNO-processed material on the surface of AG~Car. The total C+N+O abundance of AG~Car is in broad agreement with a solar composition.
 
The CNO cycle is also supposed to enhance the Na abundance in massive stars by a factor of $\sim4$ \citep{prantzos86}, which could be investigated using the wind component of the optical resonance lines of Na I 5885--5890 $\ang$. We found that Na is overabundant by a factor of $3.3 \pm 1$ in comparison to the 
solar value.

The rich emission-line spectra of AG~Car also allowed us to constrain the Si and Mg surface abundances. The former was
obtained using the ultraviolet and optical lines of \ion{Si}{2}, \ion{Si}{3}, and \ion{Si}{4}, while the latter was constrained using \ion{Mg}{2} lines at
4481, 10915, and 10952 $\ang$. For both species, the derived abundance is consistent with the solar value within 30\%.

Iron contributes significantly to the emergent spectrum of AG~Car and, together with other iron-group elements, is the main responsible for the severe line
blanketing seen in the ultraviolet spectrum. Solar abundance was initially assumed for Fe, Ni, Cr, Mn, and Co, and was found to be consistent with 
the line blanketing seen in the UV. The severe line blending with Fe lines and uncertainties in the atomic parameters make it impossible to
constrain the Ni, Cr, Mn, and Co abundance better than a factor of 2. The Fe content can be better constrained to within $\pm$30\% of the solar
value using the numerous \ion{Fe}{2} and [\ion{Fe}{2}] lines in the optical spectrum and the \ion{Fe}{2} forest in the ultraviolet.

\begin{deluxetable}{lccr}
\tablewidth{0pt}
\tabletypesize{\scriptsize}
\tablecaption{Surface chemical abundances of AG~Car \label{abund_table}}
\tablehead{\colhead{Species} & \colhead{Number fraction}  & \colhead{Mass fraction} & \colhead{Z/Z$_{\odot}$\tablenotemark{a}}}
\startdata
H  & 1.00    & $3.6\times 10^{-1}$             & $0.49 \pm 0.05$ \\
He & 0.43    & $6.2\times 10^{-1}$             & $2.5 \pm 0.2$ \\ 
C  & $5.6\times 10^{-5}$ & $2.4\times 10^{-4}$ & $0.11 \pm 0.03$ \\
N  & $1.4\times 10^{-3}$ & $7.2\times 10^{-3}$ & $11.5 \pm 3.4 $ \\
O  & $4.1\times 10^{-5}$ & $2.4\times 10^{-4}$ & $0.04 \pm 0.02$ \\
Na & $1.0\times 10^{-5}$ & $8.4\times 10^{-5}$ & $3.3 \pm 1$ \\
Mg & $7.3\times 10^{-5}$ & $6.5\times 10^{-4}$ & $1.0  \pm 0.3$\\
Al & $5.7\times 10^{-6}$ & $5.6\times 10^{-5}$ & $1.0 \pm 0.3$ \\
Si & $6.5\times 10^{-5}$ & $6.6\times 10^{-4}$ & $1.0  \pm 0.3$\\
Cr & $9.0\times 10^{-7}$ & $1.7\times 10^{-5}$ & $1.0^{+1}_{-0.5}  $\\
Mn & $4.7\times 10^{-7}$ & $9.4\times 10^{-6}$ & $1.0^{+1}_{-0.5}$\\
Fe & $6.7\times 10^{-5}$ & $1.4\times 10^{-3}$ & $1.0 \pm 0.3$ \\
Co & $1.6\times 10^{-7}$ & $3.5\times 10^{-6}$ & $1.0^{+1}_{-0.5}$\\ 
Ni & $3.4\times 10^{-6}$ & $7.3\times 10^{-5}$ & $1.0^{+1}_{-0.5}$\\
\enddata
\tablenotetext{a}{Ratio between the mass fraction of a given species in AG~Car and the respective solar mass fraction. Quoted errors indicate the range of abundances consistent with the observations. Solar abundances are from \citet{grevesse07}.}
\end{deluxetable}

\subsection{\label{jun90}Stellar and wind parameters during the 1985--1990 minimum}

We divided the data obtained from 1985--1990 into three subsets which were analyzed separately as follows. The first epoch comprised spectra from 1985 July until 1986 June, since the H$\alpha$ spectrum obtained in 1986 June showed significantly enhanced emission compared to other epochs (Fig. \ref{varhalpha90}). In the second subset, the data from 1987 January until 1990 June were averaged and analyzed as a single-epoch observation, since very little variability was present. We believe that the fundamental parameters obtained during these epochs are the least affected by time-dependent effects. The third subset consisted of data from 1990 December until 1991 January, since the \ion{He}{2} 4686 $\ang$ emission was noticeably weaker than other epochs during this minimum (Fig. \ref{varhalpha90}).

\begin{figure}
\resizebox{\hsize}{!}{\includegraphics{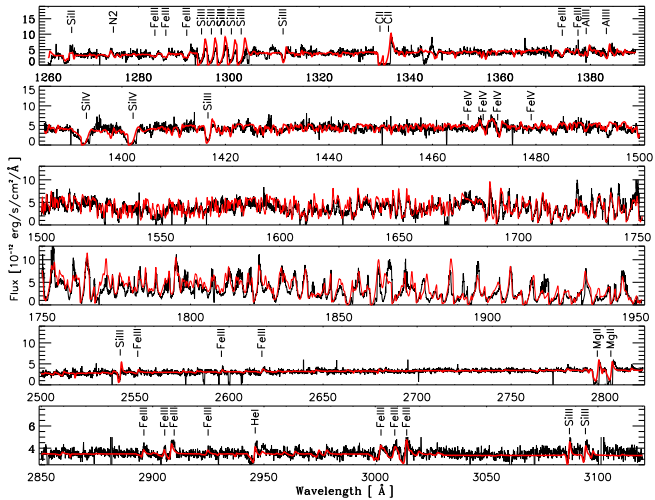}}
\caption{Comparison between the ultraviolet spectrum of AG~Car from 1989 December 23 (black line) and the best CMFGEN model for that epoch reddened using $E(B-V)=0.65$ and $R_V=3.5$ (red line) in the spectral region between $1260-3150~\ang$. The same model provides similarly good fits to the other ultraviolet spectra of AG~Car taken between 1985--1990, since there was little variability in the UV spectrum of AG~Car during these epochs \citep{leitherer94}. Line identification for the strongest spectral features is presented. The spectrum in the region between $1500-1950~\ang$ is highly blanketed and dominated by blends of \ion{Fe}{3}, \ion{Fe}{4}, and \ion{Ni}{4} lines. Notice that \ion{Fe}{2} lines are absent. The \ion{P}{3} doublet lines at 1344.32--1344.85 {\AA} were not included in 
the models.
\label{agc90a}}
\end{figure}

\begin{figure}
\resizebox{0.9\columnwidth}{!}{\includegraphics{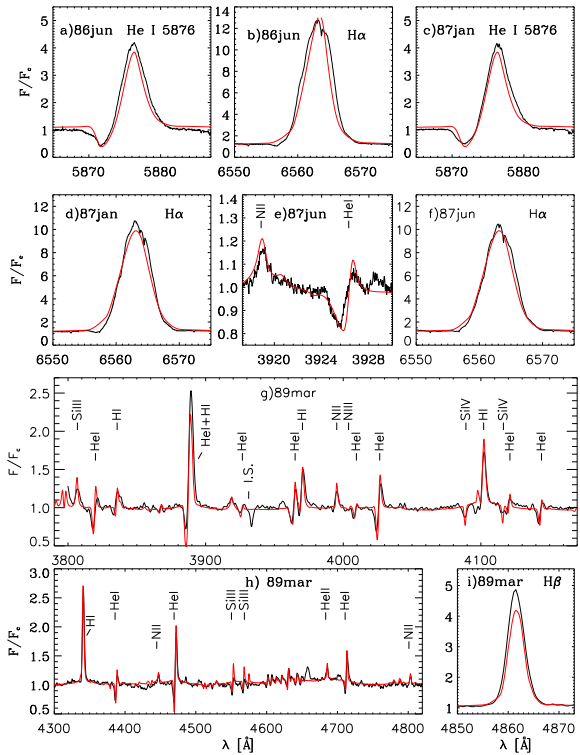}}
\caption{Optical spectra of AG~Car taken at different epochs during the 1985--1990 minimum (black line) compared with the respective best CMFGEN model for each epoch (red line; see Table \ref{agc01physpar}). Note the presence of [\ion{Fe}{3}] 4658 $\ang$ during 1989 March in panel {\it h}. \label{agc90b}}
\end{figure}

\begin{figure}
\resizebox{0.9\columnwidth}{!}{\includegraphics{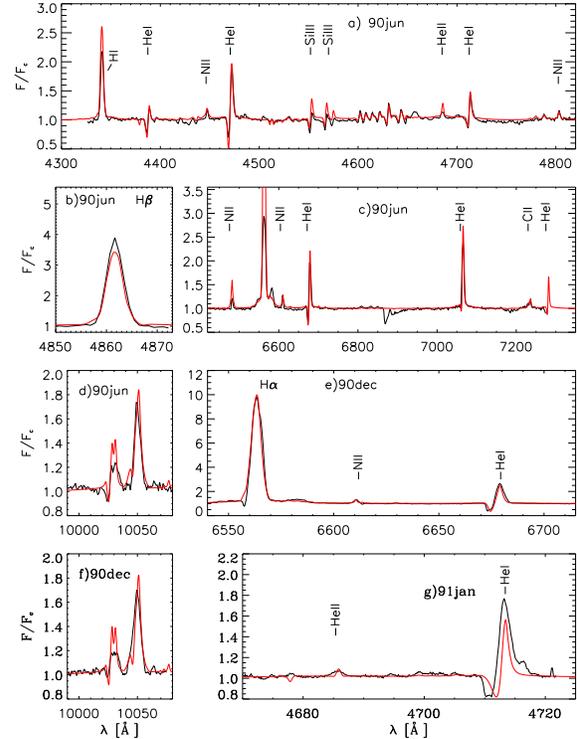}}
\caption{Optical and near-IR spectra of AG~Car taken at different epochs during the 1985--1990 minimum (black line) compared with the respective best CMFGEN model for each epoch (red line; see Table \ref{agc01physpar}). Notice in panel $g$ that \ion{He}{2} $4686~\ang$ is broader in the observations than in the CMFGEN model, which is due to the fast rotation of AG~Car (\citealt{ghd06}; Paper II). In addition, the observed \ion{He}{1} 4713 $\ang$ line profile is broader than the model in 1991 January (panel {\it g}).\label{agc90c}}
\end{figure}

Figures \ref{agc90a}, \ref{agc90b}, and \ref{agc90c} show the comparison between the best CMFGEN models obtained for the 1985--1990 minimum phase and 
the ultraviolet, optical, and near-infrared spectra of AG~Car. The models quantitatively reproduce the observed spectrum with a superb fit
in most spectral regions, in particular in the ultraviolet, where the line blending and blanketing are severe.

Table \ref{agc01physpar} presents the fundamental parameters of AG~Car during the 1985-1990 minimum, and below we discuss the diagnostics used to derive each of them.

\subsubsection{Effective temperature (\teff) and temperature at the base of the wind (\tstar)}

During this minimum, the most sensitive diagnostic for deriving $\teff$ was the ratio between \ion{He}{2} 4686 $\ang$ and \ion{He}{1} lines at 
4471 $\ang$, 4713 $\ang$, 5876 $\ang$, 6678 $\ang$, and 10024 $\ang$. Changes of a few hundred K in $\teff$ were sufficient to change the EW of \ion{He}{2} 4686 $\ang$ by a factor of 2. In addition, the ionization structure of C, N, and Si were also sensitive to $\teff$ and yielded similar
values to $\teff$. For the C ionization structure, we used the ultraviolet \ion{C}{2} 1331--1334 $\ang$, \ion{C}{3} 1175 $\ang$, and  \ion{C}{3} 1247 $\ang$ and the optical \ion{C}{2} 6578--6582 $\ang$ and \ion{C}{2} 7231--7236 $\ang$ diagnostic lines. In the case of the N ionization structure, we compared the strength of several optical \ion{N}{2} lines such as \ion{N}{2} 4601-07--30--43 $\ang$, \ion{N}{2} 6482 $\ang$, and \ion{N}{2} 6613 $\ang$ with the \ion{N}{3} triplet at 1747--1749--1751 $\ang$. The absence of \ion{N}{3} lines around 4500 $\ang$ was used to constrain the maximum value of $\teff$ during 1987--1990. The Si ionization structure was constrained through the relative strength of \ion{Si}{2}, \ion{Si}{3}, and \ion{Si}{4} lines, using lines of \ion{Si}{2} 6347 $\ang$, \ion{Si}{3} 1280--1290 $\ang$, \ion{Si}{3} 4553--4568--4575 $\ang$, \ion{Si}{4} 1393--1402 $\ang$, and \ion{Si}{4} 4088-4116 $\ang$. 

Using such a large number of relatively high-sensitive diagnostics available for $\teff$, we obtained $\teff \simeq 22,800 \pm$ 500 K during 1987--1990. There was no evidence from the 1985 July and 1986 June UV spectra that $\teff$ was higher than during 1987--1990 within the errors, and therefore we also adopted $\teff \simeq 22,800 \pm 500$ K for both epochs. Even if the \ion{He}{2} 4686 $\ang$ emission was slightly weaker in 1990 June than in 1989 March (Fig. \ref{varhalpha90}), changing $\teff$ by a few hundreds of K within the given error is sufficient to fit this line. Thus, for simplicity and due to the absence of other extremely sensitive diagnostic lines, we assume that $\teff$ was constant during 1987--1990. The smaller \ion{He}{2} 4686 $\ang$ emission during 1990 December--1991 January, when AG Car was brightening, suggests that the effective temperature was approximately 1000~K lower at that epoch ($\teff \simeq 21,500 \pm 500$ K).

Since the photosphere of AG~Car is extended, we found that $\teff$ depends on the derived value of $\mdot$ (Sect. \ref{mdotagchot90}), wind velocity law (Sect. \ref{vinfagchot}), and $M$ ($\sim 70~\msun$, Paper II). In addition, we found that the velocity and density structure below the sonic point significantly affect the value of $\teff$. In our best models, the velocity structure was iterated at depth in order to fulfill the momentum equation below the sonic point. Initial models, which assumed a fixed scale height across the extended atmosphere, yielded higher values of $\teff$ by up to $\sim1000$ K.

The value of $\tstar$ is defined in this paper as the temperature of the hydrostatic core (i.e at the base of the wind, $\rstar$) where $v=v_\mathrm{sonic}/3$ (Sect. \ref{agcmodel}), and the difference between $\tstar$ and $\teff$ of a given epoch is directly related to the extension of the atmosphere. We obtained \tstar=26,200 $\pm$ 500 K during 1987 January -- 1990 June and $\tstar \simeq 24,640 \pm 500$~K during 1990 December--1991 January. Since we assumed that $\teff$ was constant from 1985 July--1990 June, the higher mass-loss rate of AG~Car obtained from 1985 July--1986 June compared to 1987 January -- 1990 June (see Sect. \ref{mdot89}) implies that the atmosphere was more extended between 1985 July--1986 June. Therefore, $\tstar$ was slightly higher between 1985 July--1986 June  ($\tstar \simeq 26,450 \pm 500$~K) than during 1987 January -- 1990 June.

\subsubsection{Mass-loss rate and wind clumping \label{mdotagchot90}} \label{mdot89}

\begin{figure}
\resizebox{0.9\columnwidth}{!}{\includegraphics{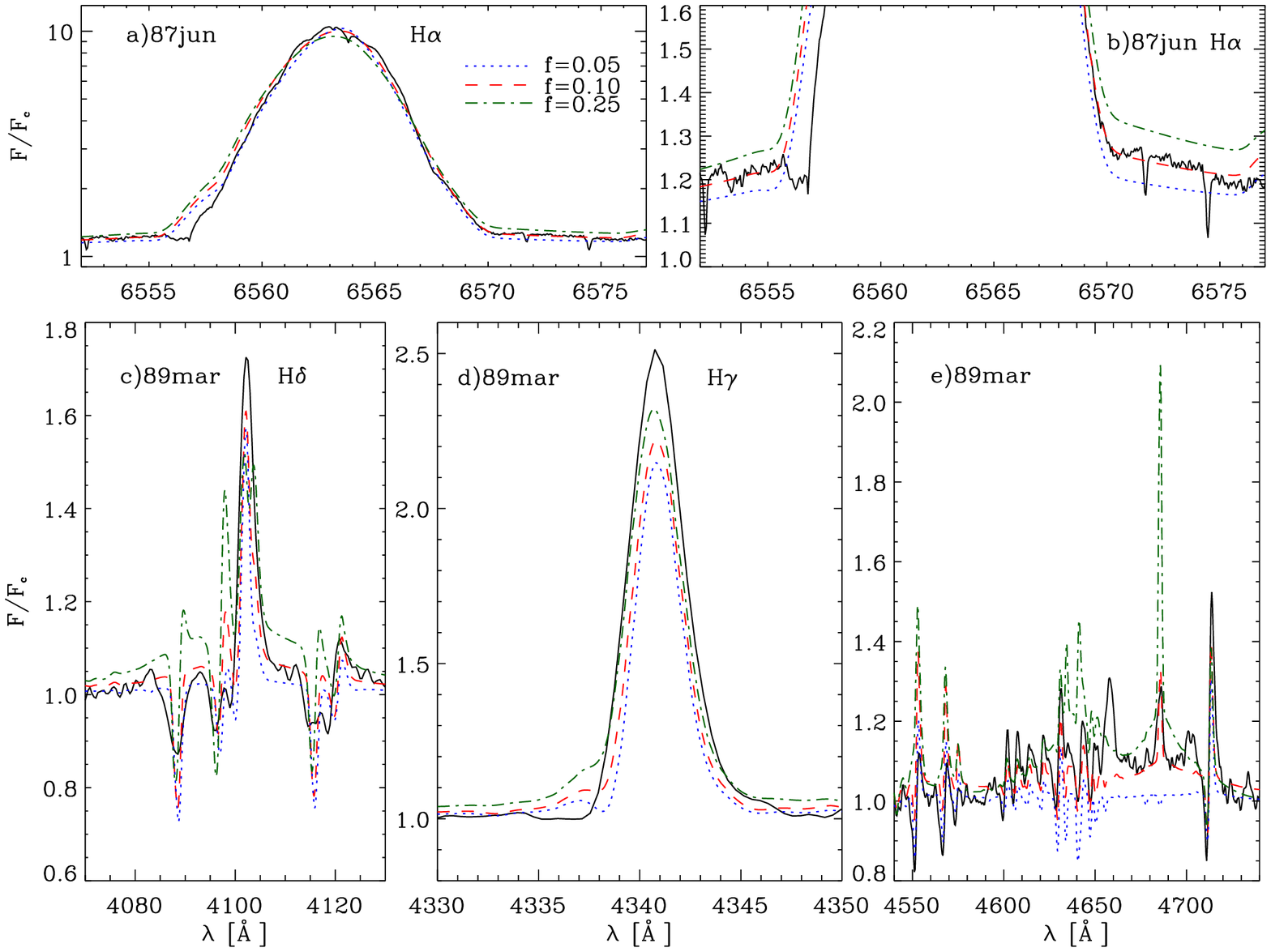}}
\caption{Optical spectra of AG~Car taken at different epochs during the 1985--1990 minimum (black line) compared with CMFGEN models assuming different volume-filling factors $f$ and with $\mdot$ scaled to produce a similar amount of H$\alpha$ emission (i.e., all models have the same $\mdot f^{-0.5}=4.7\times10^{-5}$~\msunyr). For each epoch we display models with $f=0.05$ (dotted blue line), $f=0.10$ (dashed red line), and $f=0.25$ (dot-dashed green line). Notice that the model with $f=0.25$ produces too strong electron scattering wings and increased \ion{He}{2} 4686~$\ang$ emission. Increasing the characteristic velocity at which clumps start to form ($v_\mathrm{c}$) from 20 to $70~\kms$ produces a similar effect to the \ion{He}{2} 4686~$\ang$ emission as increasing the volume-filling factor, since this line is formed very deep in the wind. Note in panel $c$ that the \ion{Si}{4} $4088-4116~\ang$ absorptions are broader in the observations than in the CMFGEN model because rotational broadening was not included (\citealt{ghd06}; Paper II). \label{clump}}
\end{figure}

The mass-loss rate was obtained by reproducing the intensity of the strongest lines in the observed spectrum of AG~Car: namely, those
of hydrogen, helium, and nitrogen. Using those diagnostics we obtained $\dot{M}=1.5 \times 10^{-5}~\msunyr$ during 1987--1990. Spectra from 1990 December--1991 January are also consistent with the same $\mdot$. Figure \ref{clump} illustrates the effects of wind clumping on selected spectral lines of AG Car, presenting CMFGEN models assuming different volume-filling factors but with the same value of $\mdot f^{-0.5}=4.7\times10^{-5}$~\msunyr and, thus, equally reproducing the amount of H$\alpha$ emission. We derived a volume-filling factor of $f=0.1$ based on the strength of the electron-scattering wings of H$\alpha$, H$\beta$, H$\gamma$, H$\delta$, and \ion{He}{1} 6678 $\ang$.

Similar to what has been found for O-stars \citep{hillier03,bouret03,bouret05}, we found that clumps must be created very deep in the wind, close to the sonic point (i.e. $v_\mathrm{c}\simeq 20~\kms $), in order to reproduce the weak electron-scattering wings of \ion{He}{2} 4686 $\ang$. The presence of clumps very close to the stellar surface is in line with the latest polarimetric results by \citet{davies05,davies06a,davies07,davies08}.

The statistical error in $\mdot$ when using a single spectral line, such as H$\alpha$, as diagnostic is as small as 5\%. However, taking into account that multiple lines need to be fitted, that additional parameters can also change the line strength and have their own errors, and the assumptions in the model (prescribed velocity law, steady-state, spherical-symmetric wind, etc.), we estimate that the uncertainty in $\mdot$ amounts to 30\%. We found that models assuming either $f \leq 0.05$ or $f \geq 0.25$ do not provide reasonable fits to the electron-scattering wings present in the AG~Car spectrum (Fig. \ref{clump}).    

As shown in Sect. \ref{evolhot90}, moderate spectroscopic variability is seen during the minimum only in the H$\alpha$ profile obtained in 1986 June, which was $\sim 20\%$ stronger than during the rest of the minimum (1987--1990). To investigate the nature of the changes seen in H$\alpha$ and explain the behavior of AG~Car during 1985--1990, we explored the parameter space around the best CMFGEN radiative transfer model obtained for 1987--1990 (see Table \ref{agc01physpar}). This model spectrum does not have a P-Cygni absorption component in H$\alpha$ and provides a nice fit to the H$\alpha$ profile from 1987 June -- 1990 June. 

We found that the H$\alpha$ emission in 1986 June can be explained by an increase of $\sim30\%$ in $\mdot$ compared to the value obtained for the rest of the minimum  ($\dot{M}=1.5 \times 10^{-5}$~\msunyr, see Sect. \ref{mdotagchot90}), corresponding to $\dot{M}=1.9 \times 10^{-5}$~\msunyr. However, increasing $\mdot$ by 30\% is not sufficient to change the H ionization structure and does not produce a weak P-Cygni absorption component. Furthermore, the change in $\mdot$ by 30\% in the model produces a modest 1\% increase in the $y$-band flux, or 0.01 mag. Interestingly, such an amount of photometric variability induced by the change in $\mdot$ is compatible with the microvariability detected in AG~Car during minimum \citep{vg88,vg90}.

As has been suggested in order to explain the spectroscopic changes seen in the LBVs He 3-519 \citep{crowther97} and P Cygni \citep{pauldrach90,najarro97}, we also found that the H ionization structure of AG~Car could be affected by changes in $\mdot$, $\rstar$, and $\lstar$. However, in the parameter regime found during 1985--1990, we determined that even if $\mdot$ is increased by a relatively large factor of 4, the model spectrum does not show a weak P-Cygni absorption component, as seen in the observations. Therefore, a decrease in $\teff$ and/or $\lstar$ is required in order to fit the weak P-Cygni absorption in the 1986 June spectrum. The decrease in $\teff$ is not supported by the ultraviolet spectrum taken in 1985--1990, while the change in $\lstar$ would produce, according to our models, significant photometric changes of at least $\sim 0.5$ mag, which is not observed during 1985--1990. 

\subsubsection{Wind terminal velocity and velocity law \label{vinfagchot}}

We used the maximum velocity of the saturated P-Cygni absorption component of the UV resonance lines\footnote{The so-called $v_\mathrm{black}$, see \citet{prinja90}.} of \ion{C}{2} 1334--1335 $\ang$, \ion{Si}{4} 1393--1402 $\ang$, and Mg II 2798--2810 $\ang$ as a diagnostic to derive the wind terminal velocity. In the case of \ion{C}{2} 1334--1335 $\ang$ both lines are blended, and only the blue component was used. The determination of $\vinf$ from UV resonance lines is very weakly dependent on the assumed velocity law. Since the UV resonance lines essentially have no variability during minimum \citep{leitherer94}, all epochs between 1985--1989 yield $\vinf=300 \pm 30~\kms$.
 
Additional constraints on the value of $\vinf$ can be obtained from the FWHM of the emission component of the strongest optical lines, such as H$\alpha$, H$\beta$, H$\gamma$, and \ion{He}{1} lines, and from the P-Cygni absorption profile of the \ion{He}{1} lines. Using these diagnostics, we derived $\vinf=270 \pm 50~\kms$, which is slightly lower but still in line with the value obtained from the UV resonance lines. The optical lines provided only a lower limit to $\vinf$, since they were much more sensitive to the adopted velocity law and to the density structure of the wind than the UV resonance lines, which explains the different results comparing optical and UV analyses. Using only the optical lines, lower values of $\vinf$ were obtained when low values of the wind acceleration parameter $\beta$ were used. In particular, a normal velocity law  with $\beta=1$ would require $\vinf$ as low as $230~\kms$, while higher values of $\beta$ such as 3--4 would require a value of $\vinf$ closer to that obtained from the UV lines. Thus, ultimately, we favor the use of the value obtained from the UV lines, $\vinf=300 \pm 30~\kms$, and that $\beta$ is roughly $3$. A high value of $\beta$ is supported by several other diagnostics. Low values of beta did not simultaneously provide reasonable fits to UV resonance lines and to the optical \ion{H}{1} and \ion{He}{1} line profiles, in particular to their absorption profile. In addition, the ratio between H$\alpha$ and the other Balmer lines such as H$\beta$ and H$\gamma$ required $\beta \geq 3$. Values of $\beta < 3$ also yielded too much radiative acceleration beyond the sonic point (Paper II).

Evidence of an extended acceleration zone in the outer wind is suggested by the wind hydrodynamics: a velocity law using a single value of $\beta$ yielded too much radiative acceleration for distances greater than $\sim 10^2~\rstar$ (Paper II). However, we chose to use a usual beta-type law in order to not introduce extra free parameters in the analysis.

Line-driven winds are intrinsically unstable, and turbulent motions are expected to arise \citep{owocki88,feldmeier95}. Such a turbulent velocity field produces a non-saturated absorption wing extending to velocities higher than $\vinf$ on most UV resonance lines in the AG~Car spectrum. In addition, a redshift of the emission components of the stronger \ion{H}{1}, \ion{He}{1}, and \ion{Fe}{2} lines is also produced due to the turbulent velocity field \citep{hillier87,hillier89,catala84}. In CMFGEN, it is possible to account for a microturbulent velocity $v_\mathrm{turb}$, which was set to $20~\kms$. The use of higher values of v$_{turb}$ causes a redshift of the emission component of the stronger \ion{He}{1} and hydrogen lines, which is not seen in the observed spectrum. This also explains the shift of $\sim+10 \kms$ of the H$\alpha$ emission component, relative to the systemic velocity, measured by \citet{stahl01} (the same behavior is observed in our data). We obtained satisfactory fits to the extended, non-saturated absorption wing using v$_{turb}=20~\kms$.

\citet{leitherer94} noticed that during minimum the UV resonance lines of \ion{Si}{4} 1393--1402 $\ang$ have an extended, non-saturated
absorption wing with velocities up to $\sim -700~\kms$. Those authors interpreted this component either as evidence of a fast polar 
outflow of AG~Car or as being caused due to blending with Fe lines (following \citealt{hubeny85}). Our models predict severe blending due to \ion{Fe}{3} and \ion{Fe}{4} around both \ion{Si}{4} lines, which can mimic a high-velocity absorption wing. Therefore, we do not support the idea that the blueshifted, high-velocity component seen in the \ion{Si}{4} lines are actually due to a fast polar wind; instead, we support that they are caused by \ion{Fe}{3} and \ion{Fe}{4} absorption lines. The high projected rotational velocity inferred for AG~Car by \citet{ghd06}, and the inference that we are seeing AG~Car close to edge-on, provides further evidence that the high-velocity absorption does not come from the polar wind.

\subsection{\label{2001-03}Stellar and wind parameters during the 2000--2003 minimum}

Unlike the previous minimum, the observations of AG~Car between 2000--2003 show significant temporal variability. The data were divided in four datasets consisting of data between 2000 July--2001 June, 2002 March--2002 July, 2002 November, and 2003 January. There was little variability during the time intervals of 2000 July--2001 June and 2002 March--2002 July; therefore, the average spectrum was used for those intervals. As a consequence of the larger spectral coverage and higher spectral resolution, it was possible to use a larger number of optical diagnostic lines during 2000--2003 in comparison to the previous minimum. The spectra taken before 2000 July and after 2003 January indicate $\teff \leq 13,000 \mathrm{K}$, constraining the duration of this minimum to at most 2.5
years. The maximum $\teff$ was only achieved during $\sim$1--1.5 years (Table \ref{agc01physpar}).

The absence of ultraviolet observations did not affect the determination of the stellar and wind parameters of AG~Car since a much larger spectral coverage in the optical and near-IR was available compared to the 1985--1990 minimum, providing enough diagnostics. In addition, the main
parameter obtained from ultraviolet lines, namely the wind terminal velocity, was determined based on the numerous P-Cygni
absorption components of \ion{He}{1}, \ion{N}{2}, \ion{Fe}{2}, and hydrogen Balmer and Paschen lines. For the data taken in 2001 May, it was also possible to
constrain $\vinf$ using \ion{C}{3} 1173 $\ang$ from the FUSE spectrum, yielding a similar value of $\vinf$ as the one found from the optical
lines.

The optical diagnostic lines used to obtain the stellar and wind parameters are similar to those described in
Sect.~\ref{jun90} and are just outlined here. The value of $\mdot$ was obtained in order to reproduce the strength of the H, He, N, O, Mg, Si and Fe lines, while the volume-filling factor $f$ was constrained using the intensity of the electron-scattering wings of the strongest H, \ion{He}{1}, and \ion{Fe}{2} lines.

The value of $\teff$ was determined based on the ionization structure of C, N, O, and
Si, and on the relative strength between H and  \ion{He}{1} lines, assuming the He abundance obtained for the previous minimum. The presence of very weak \ion{He}{2} 4686 $\ang$ emission was an additional constraint for the maximum value of \teff. The N ionization structure was constrained by comparing the strength of numerous spectral lines\footnote{We refer to the atlas of \citet{stahl93} for a comprehensive identification of the spectral lines found in the parameter regime investigated in the present paper.} of \ion{N}{1} (e.g., 7442, 7468, 8184-88, 8680-83-86, 8703-11-18 $\ang$), \ion{N}{2} (e.g., 4607--4630--4643, 6482 $\ang$), and \ion{N}{3} $4097~\ang$. Likewise, several \ion{O}{1} (e.g. $8446~\ang$) and many weak \ion{O}{2} lines (e.g., 3973, 4349, 4070-72, 4317-19-49, $4649~\ang$) present in the optical spectrum of AG Car were used to derive $\teff$. Assuming the C abundance obtained for the previous minimum, the C ionization structure could be constrained based on the strength of several \ion{C}{2} lines, such as those at $6578-82~\ang$ and $7231-36-37~\ang$. Optical spectral lines of \ion{Si}{2}, \ion{Si}{3}, and \ion{Si}{4} were also used as diagnostics of $\teff$. The values of $\teff$ provided by different diagnostics were consistent within $\sim500$~K. 

The strength of \ion{Fe}{3} and some \ion{Fe}{2} lines were well reproduced by the CMFGEN models, supporting the value obtained for $\teff$ and $\mdot$. Note that a few \ion{Fe}{2} lines, in particular those formed in the outer part of the wind such as \ion{Fe}{2} $5169~\ang$, were stronger in the models than in the observations. We suspect that these lines are affected by time-dependent effects.

Figures \ref{agc01a}, \ref{agc01b}, and \ref{agc01c} present the observed AG~Car spectra from 2001 April, 2002 March, and 2003 January, respectively, 
and the corresponding best CMFGEN model for each epoch. In general, a reasonably good fit to the observed spectrum could be obtained. However, in the 2001 April data a weak excess emission on top of the P-Cygni profile of {\sc H} and \ion{He}{1} is present approximately between +130 and +250 $\kms$, and is not reproduced by the models. Such an excess emission could be caused by a latitude-dependent wind or by a velocity law different than the usual beta-type law assumed in this work. The values obtained for \tstar, \teff, \vinf, \mdot, and $f$ are listed in Table \ref{agc01physpar}.

\setcounter{figure}{9}
\begin{figure*}
\figurenum{9a}
\resizebox{\hsize}{!}{\includegraphics{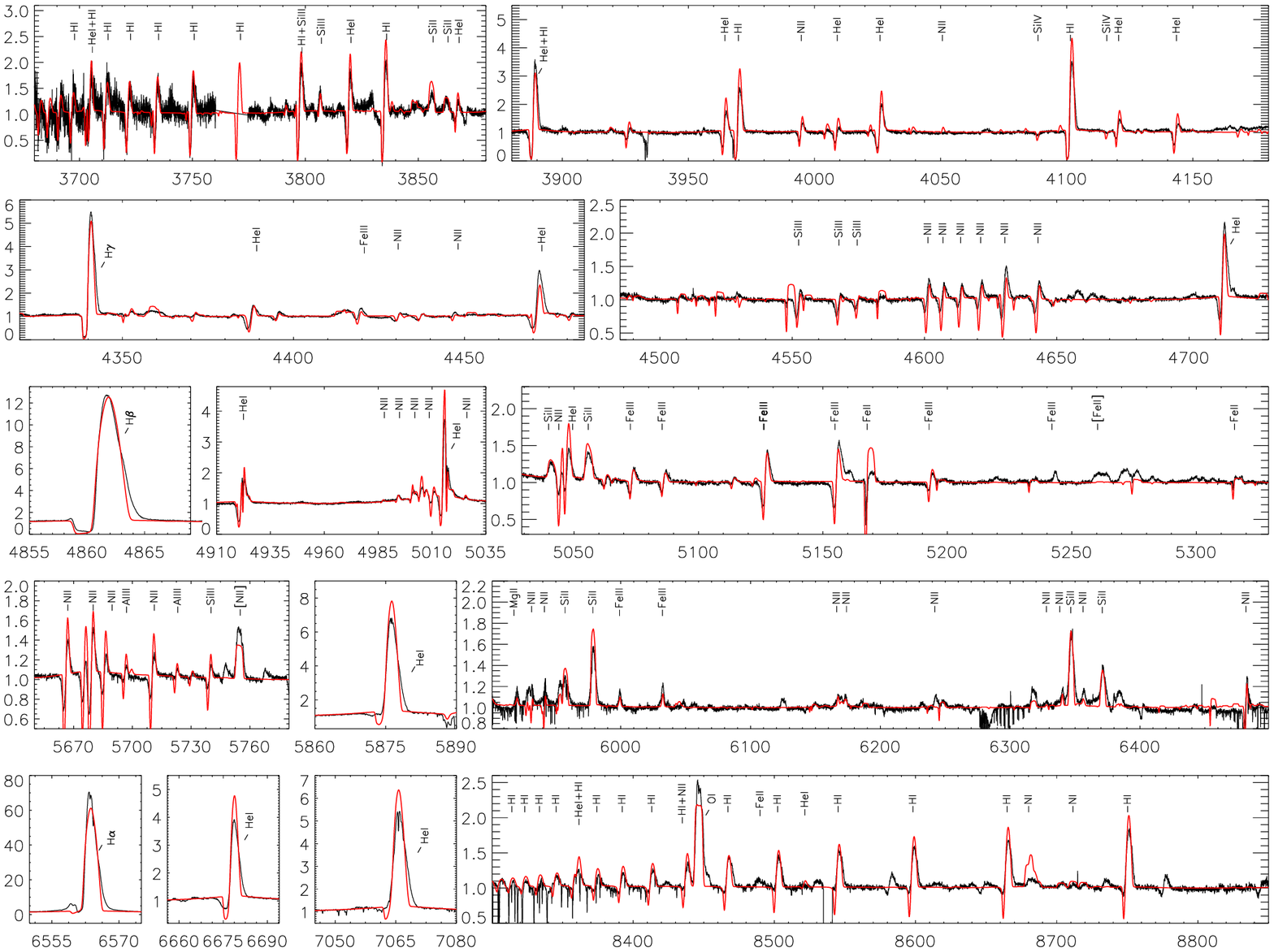}}
\caption{Comparison between the observed spectrum of AG~Car from 2001 April (black line) and the best model obtained with CMFGEN (red) in the
spectral region between 3680--11000 $\ang$. For all plots, the x-axis is the wavelength in Angstroms, while the y-axis represents the continuum-normalized flux. The identification of the stronger spectral lines is also shown {\it [see the electronic version of the journal for fits to 2002 and 2003 data].}\label{agc01a}}
\end{figure*}

\begin{figure*}
\figurenum{9b}
\resizebox{\hsize}{!}{\includegraphics{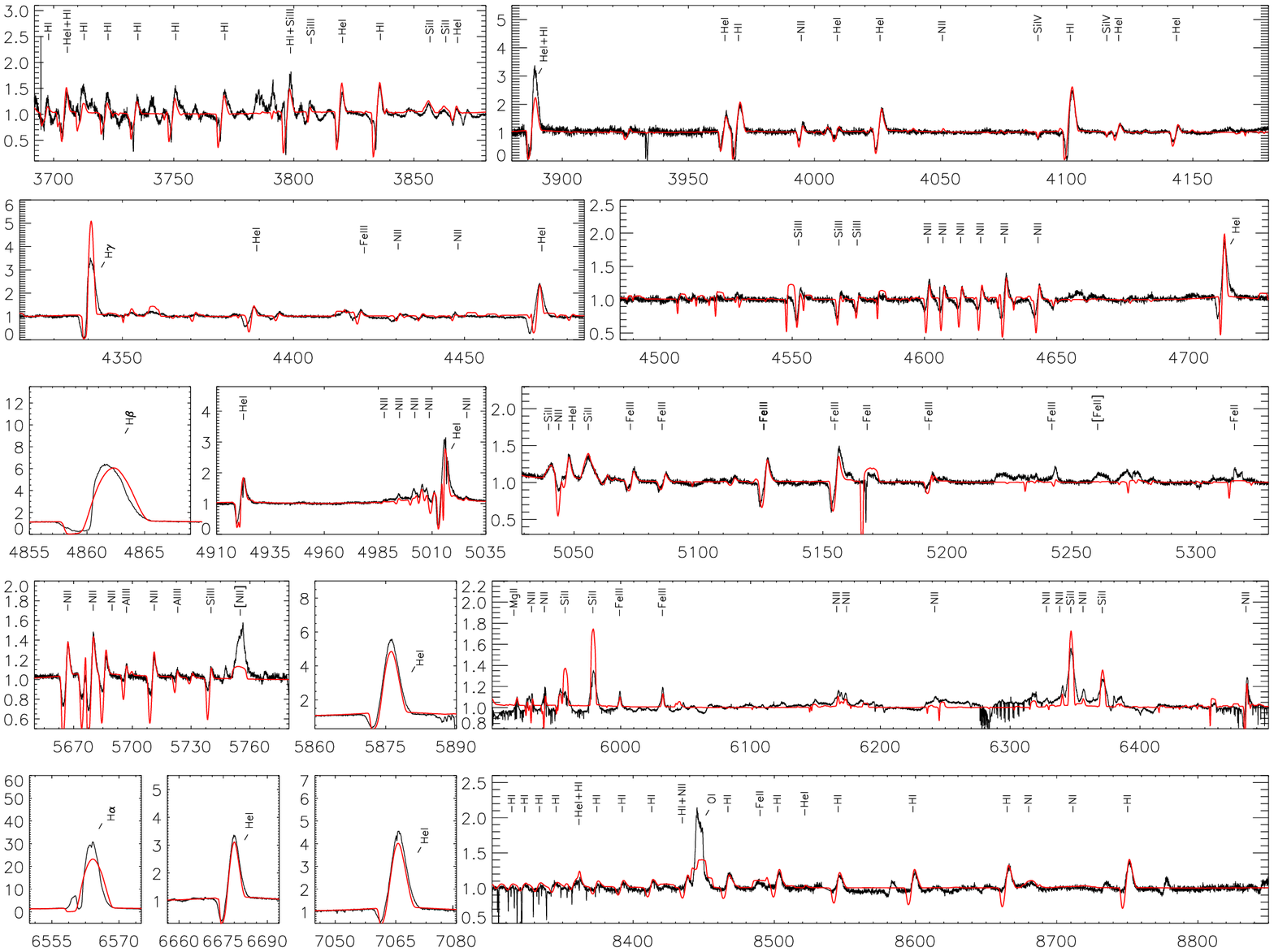}}
\caption{Similar to Fig. \ref{agc01a}, but for data taken in 2002 March. \label{agc01b}}
\end{figure*}

\begin{figure*}
\figurenum{9c}
\resizebox{\hsize}{!}{\includegraphics{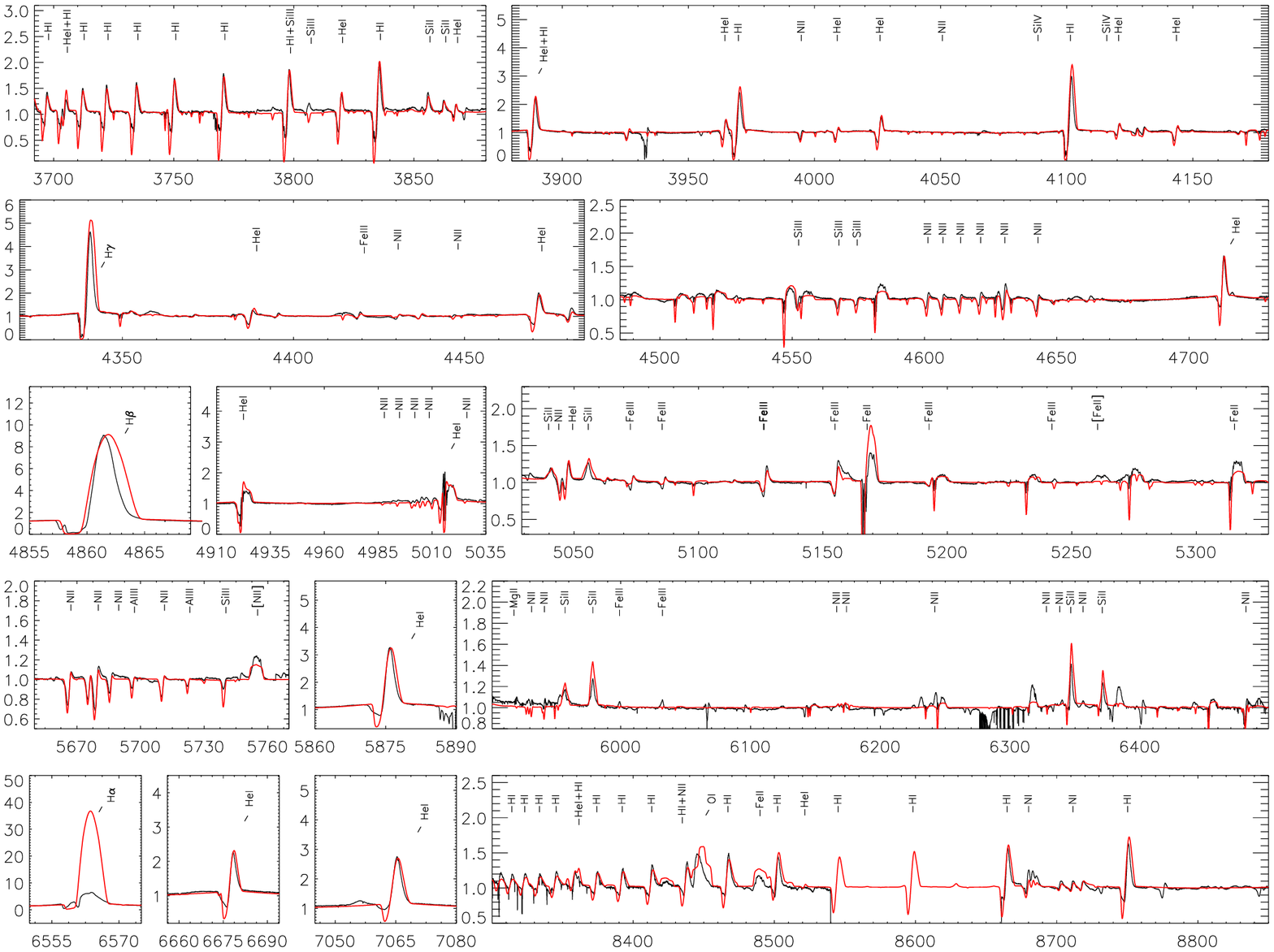}}
\caption{Similar to Fig. \ref{agc01a}, but for data taken in 2003 January. Note that H$\alpha$ is saturated and was not used as a diagnostic. \label{agc01c}}
\end{figure*}

\subsection{The {\em FUSE} spectrum of AG~Car and the extreme line blanketing in the far ultraviolet}

Since the launching of the Far Ultraviolet Spectroscopic Explorer (FUSE) satellite, high-resolution far-ultraviolet spectroscopy of OB and WR stars has provided invaluable constraints on their fundamental properties, since a significant fraction of the stellar light is emitted in this spectral region and important diagnostic lines fall in the far ultraviolet. For a recent review on the accomplishments of the far-UV research of massive stars, see \citet{crowther06b}.

\begin{figure*}
\resizebox{\hsize}{!}{\includegraphics{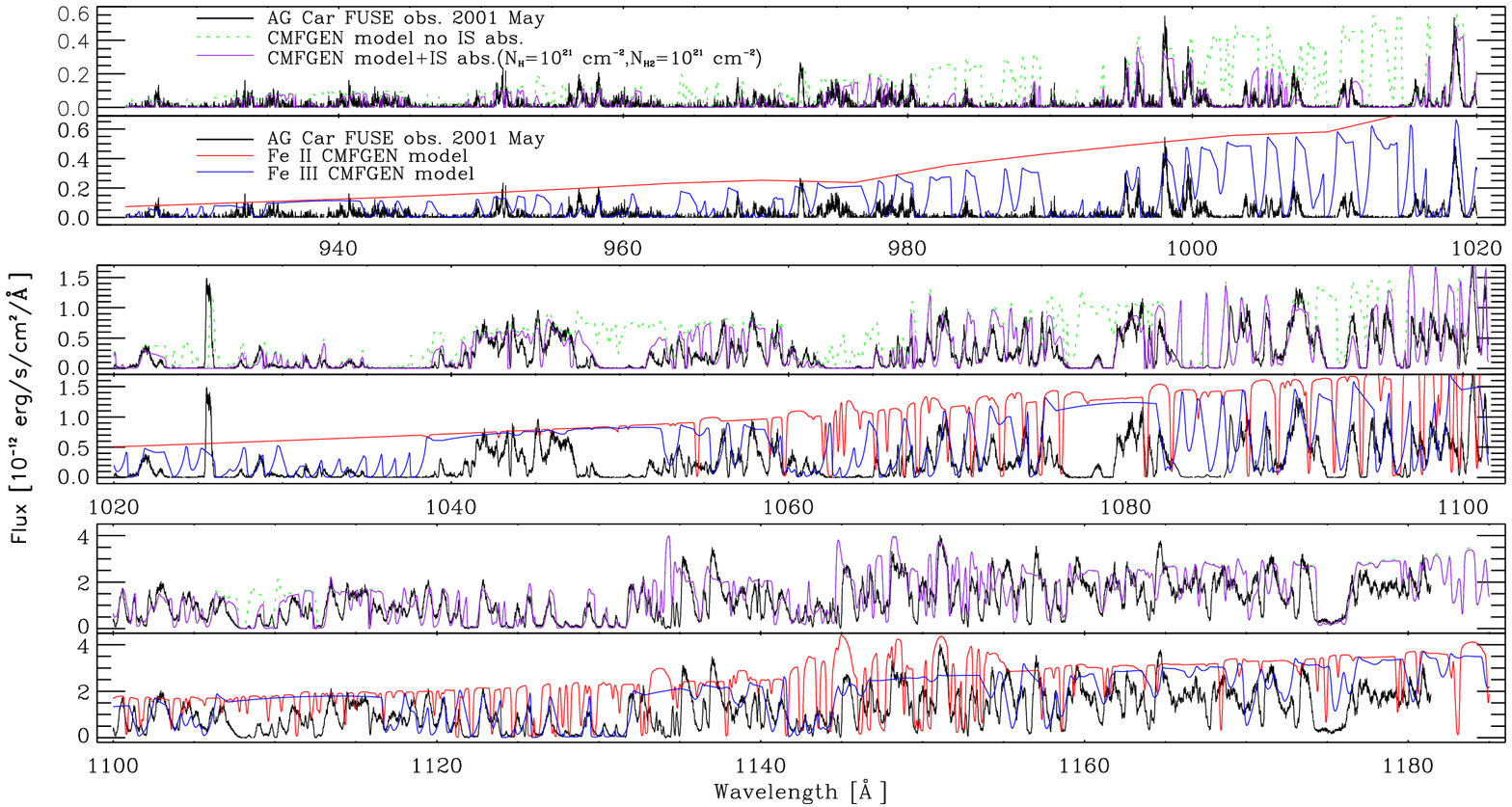}}
\caption{\label{agcfuse} Far-ultraviolet spectrum of AG~Car observed with FUSE in 2001 May (black solid line) compared with the best CMFGEN model for this epoch (green dotted line), reddened using $E(B-V)=0.65$ and $R_V=3.5$. The CMFGEN model was scaled in flux in order to better match the continuum level. The best model, taking into account an interstellar absorption spectrum due to \ion{H}{1} with $\log[N(\mathrm{H})\,\mathrm{cm}^{2}]=21.0$ and H$_2$ with $\log[N(\mathrm{H}_2)\,\mathrm{cm}^{2}]=21.0$, is shown in order to infer the influence of those transitions on the morphology of the far-ultraviolet spectrum of AG~Car (purple solid line). For each wavelength range, the bottom panel shows the best CMFGEN model spectrum computed including continuum emission and bound-bound transitions due to only \ion{Fe}{2} (red solid line) or only \ion{Fe}{3} (blue solid line), in order to facilitate line identification. These ions are the main contributors to the line blanketing in the far-UV and illustrate how well CMFGEN can handle the effects caused by the extreme blending due to tens of spectral lines.  }
\end{figure*}

Figure \ref{agcfuse} shows the comparison between the archival FUSE spectrum of AG~Car obtained during 2001 May and the best CMFGEN model for that epoch reddened using $E(B-V)=0.65$ and $R_V=3.5$ (see Sect. \ref{agcdistred}). The CMFGEN model spectrum matches reasonably well the observations, and the FUSE spectrum is consistent with the presence of a dense wind, indicating $\mdot$ in the range $3-4 \times 10^{-5}~\msunyr$. The relative strength of \ion{Fe}{2} and \ion{Fe}{3} lines is the main diagnostic for $\teff$, and agrees with the value derived from the optical/near-IR analysis. The many \ion{Fe}{2} and \ion{Fe}{3} transitions provide a good constraint on $\vinf$, as does \ion{C}{3} $1173~\ang$: all these diagnostics indicate a very low wind terminal velocity during 2001 May ($\vinf\simeq90-130~\kms$). 

It can readily be noticed in Fig. \ref{agcfuse} that line blanketing is extreme in the far-UV, and the determination of the continuum level with confidence is impossible from the observations alone. Besides \ion{C}{3} $1173~\ang$, the identification of other spectral features is hampered by the severe line overlapping present in the far-UV. In order to determine which species are responsible for a given spectral feature, we computed the observed spectrum due to a given ion using {\sc CMF\_FLUX} \citep{bh05}. We found that most of the observed spectral features in the far-UV are actually a blend of line transitions, even in the case of \ion{C}{3} 1173~$\ang$, which is contaminated by \ion{Fe}{2} and \ion{Fe}{3} lines (Fig. \ref{agcfuse}). We present in Fig. \ref{agcfuse} the separate contributions of lines of a given ion to the observed spectrum of AG~Car. Most of the line blanketing in the FUSE range is provided by \ion{Fe}{3}, while the contribution of other species are localized to a certain spectral range. In the 2001 spectrum of AG~Car, most of the \ion{Fe}{2} blanketing in the far-UV is in the spectral region 1050--1180 $\ang$; \ion{Cr}{3} contributes at 1000--1070 $\ang$,  \ion{Co}{3} at 920--965 $\ang$, and \ion{Ni}{3} around 950--980 $\ang$.

As has been seen in Galactic OB-type stars \citep{pellerin02}, the interstellar absorption due to H and H$_2$ significantly affects the observed spectrum below 1100 $\ang$. A direct visualization of the influence of the interstellar absorption on the spectrum can be seen by comparing similar stars in the Galaxy \citep{pellerin02} and the Magellanic Clouds \citep{walborn02}. In order to illustrate this effect for AG~Car we also display on Fig. \ref{agcfuse} the model spectrum taking into account the interstellar absorption by H and H$_2$, using a procedure described by \citet{herald01}, and assuming $\log[N(\mathrm{H})\, \mathrm{cm}^{2}]=21.0$ and $\log[N(\mathrm{H}_2)\,\mathrm{cm}^{2}]=21.0$. It is beyond the scope of this work to provide a detailed analysis of the interstellar spectrum towards AG~Car.

\subsection{\label{agcdistred} Luminosity, bolometric correction, and reddening}

\begin{figure*}
\resizebox{\hsize}{!}{\includegraphics{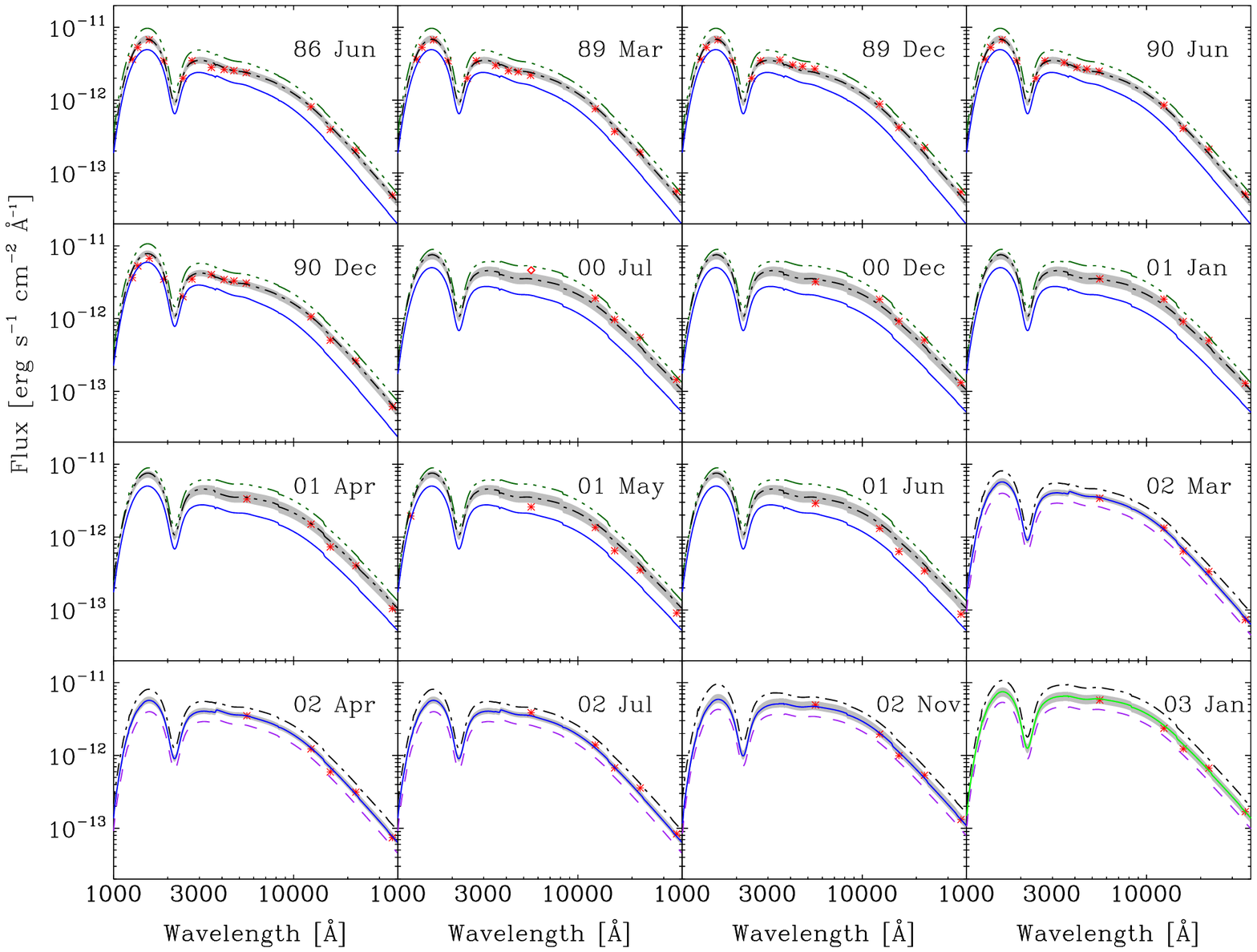}}
\caption{\label{agcflux} Comparison between the observed spectral energy distribution of AG~Car obtained from photoelectric data during minimum epochs (red asterisks) and CMFGEN models with different bolometric luminosities. For 2000 July we display the visual magnitude (red open diamond). The panels are ordered chronologically, and in decreasing effective temperature, from left to right, top to bottom. For each panel we show selected models with $\lstar=2.2\times10^{6}~\lsun$ (dark green dash-triple dotted line), $\lstar=1.5\times10^{6}~\lsun$ (black dash-dotted line), $\lstar=1.1\times10^{6}~\lsun$ (light green solid line), $\lstar=1.0\times10^{6}~\lsun$ (blue solid line), and $\lstar=0.7\times10^{6}~\lsun$ (purple dashed line). For a given epoch, models with different $\lstar$ had their $\rstar$ and $\mdot$ scaled in order to still fit the observed spectral lines. The gray region corresponds to the uncertainty in the model flux which comes from the luminosity. In all panels, the CMFGEN models were reddened using the Galactic extinction law of  \citet{fitzpatrick99} with $R_V=3.5$ and $E(B-V)=0.65$. For each epoch, there is superb agreement between the best model and the observed  data using the same reddening law, although different luminosities are needed. Note the trend towards lower luminosity as the effective temperature (radius) decreases (increases) during 2001--2003. The uncertainty in the observed photometry is smaller than the size of the symbols in the UV, optical, and in the near-IR.}
\end{figure*}

The luminosity of AG~Car during the epochs presented in this paper was constrained by comparing the observed flux of each epoch (Table \ref{obsirphot1}) with the reddened flux predicted by the CMFGEN model (Fig. \ref{agcflux}). For the minimum phase of 1985--1990, we used observations from the ultraviolet (1200 $\ang$) to the near-infrared ($L$ band), while for the 2000--2003 minimum, we had available photometry in the $V$, $J$, $H$, $K$, and $L$ bands. A sole far-ultraviolet flux-calibrated spectrum obtained in 2001 May was additionally used for this epoch. Note that no AAVSO observations were used to derive a V-band flux due to their relatively low accuracy compared to LTPV and ASAS-3 photoelectric observations.

We obtained a bolometric luminosity for AG~Car of $\lstar=1.5\times10^{6}~\lsun$ during 1985--1990 and 2000--2001, which is listed in Table 
\ref{agc01physpar} together with the other wind and stellar parameters. However, a lower luminosity of $\lstar=1.0\times10^{6}~\lsun$ and $\lstar=1.1\times10^{6}~\lsun$ was derived for the 2002 March--July and 2003 January observations, respectively. Figure \ref{agcflux} shows a clear trend towards lower luminosity during the end of the 2000--2003 minimum. Excluding errors in the distance, the typical error in $\lstar$ solely from the spectroscopic analysis is around 10\% (gray region in Fig. \ref{agcflux}).

Because of our limited spectroscopic and photometric sample, the scope of this paper is to describe the long-term behavior of the stellar parameters of AG Car. Therefore, we did not attempt to obtain a different bolometric luminosity for epochs relatively close in time ($\Delta t \sim 3 - 6$ months), when little spectroscopic variability was seen. This was the case, for instance, for 1989 March--1989 December, 2001 April--2001 June, and 2002 March--2002 July. Although AG Car clearly presented photometric variability during these time periods, we assumed a constant bolometric luminosity for each of these periods and derived an average bolometric luminosity which, within the errors, could describe the SED during the whole time interval. Note that, according to Fig. \ref{agcflux}, such an assumption does not change the main result of this Section, i.e., a lower bolometric luminosity as the star is moving towards maximum.

A much larger amount of spectroscopic and photometric data with a higher time sampling, together with a careful treatment of time-dependent effects in the radiative transfer model, are needed to study the behavior of the stellar parameters of LBVs on timescales of the order of a couple of months. Nevertheless, if the bolometric luminosity is variable on these short timescales, its amplitude is likely much smaller ($\Delta\log [\lstar/\lsun] \sim0.02$ dex) than the amplitude of the long-term variability ($\Delta\log [\lstar/\lsun] \sim0.17$ dex). Therefore, small fluctuations of $\pm 0.02$ dex in $\log (\lstar/\lsun)$ may occur within each range of epochs shown Table \ref{obsirphot1}, but such variations are within the errors of our modeling.

Since we derived $\lstar$ based on the comparison between the observed and model spectral energy distributions, different values of $\lstar$ could in principle be obtained by changing $\mdot$ and $\teff$. However, unrealistic changes in these parameters are required in order to reach the same value of $\lstar$ for all the different minimum epochs; for instance, $\teff$ would need to be reduced/increased by more than 2000~K for a given epoch, which is not supported by the line fits, in order to change the derived value of $\lstar$ by 30\%. A similar drastic change in $\mdot$ by more than a factor of 2, which again is not supported by the line fits, would be required in order to change the continuum emission and thus fit the data with the same $\lstar$ for all epochs. Even then, very peculiar reddening laws, which would be different for each epoch, would be required in order to reproduce the shape of the SED.

Therefore, the decrease in the bolometric luminosity obtained during the end of the minimum is robust, implying that {\it the bolometric luminosity changes during the S-Dor cycles of AG~Car.} The similar maximum value of $\lstar$ derived for AG~Car, for both minimum phases analyzed in this work, is indicative that this is the maximum luminosity of AG~Car during minimum. On the other hand, the change in bolometric luminosity is contrary to the currently widely accepted paradigm of S-Dor variability at constant $\lstar$. A detailed analysis of the changes of $\lstar$ during the full S-Dor cycle will be presented in Paper III.

The bolometric correction ($BC$) in the $V$-band for each epoch was also obtained from our modeling and is presented in Table \ref{agc01physpar}. As a consequence of the different stellar parameters, the $BC$ of AG Car is also different in consecutive minima, ranging from $\simeq -2.50$ mag (1985--1990) to -2.15 mag (2000 July--2001 June). Further change in $BC$ occurs when AG Car is moving towards maximum ($BC\simeq -1.68$ mag during 2002 March--July and $BC\simeq - 1.22$ mag in 2003 January).  

While it is well known that the bolometric correction of stars which have a negligible wind (such as the Sun and OB stars) is dependent on $\teff$ and $\logg$, we found 
that the bolometric correction of LBVs, because of the atmospheric extension, is dependent not only on $\teff$ and $\logg$ but also on the wind density (and thus on $\mdot$ and/or $\vinf$). Since $\teff$ also depends on the wind density, LBVs with different stellar parameters can have a similar value of $\teff$ but significantly different BCs; this is exactly the case for AG Car during 2001 April and 2002 March (see Table \ref{agc01physpar}). While $\teff$ was close to $\sim17,000$ K during both epochs, the star was very different at depth, as can be readily evidenced by the very different temperatures at the hydrostatic radius in 2001 April ($\tstar \simeq 21,900$ K) and 2002 March ($\tstar \simeq 18,700$ K). Therefore, the $BCs$ were different, and the effective temperature was similar just by coincidence. A complete analysis of the bolometric correction of LBVs over a broad range of physical parameters is beyond the scope of this work, and will be presented elsewhere. 

Together with $\lstar$, the reddening law towards AG~Car was obtained through the comparison of the observed flux and CMFGEN models reddened
according to the law of \citet{fitzpatrick99}. The best fits were obtained using $E(B-V)=0.65\pm 0.01$ and $R_V=3.5 \pm 0.1$,
and are shown in Figure \ref{agcflux}. Taking the uncertainties into account, it was possible to fit the data from both minima with
the same reddening law, corroborating the physical parameters obtained for both epochs. The value of $E(B-V)=0.65$ is only slightly
higher than previous determinations ($E(B-V)=0.63$) obtained using simple models and assuming constant bolometric luminosity \citep{humphreys89,shore96}.

Although the value of $R_V=3.5$ obtained in this work is higher than the standard value due to diffuse dust ($R_V=3.1$), it is lower  
than $R_V=3.9$, as suggested by \citet{sl94} and based on the analysis of the polarimetric data of AG~Car. Our results are compatible with the presence of large grains ($\sim 1 \mu$m) in the circumstellar environment of AG~Car \citep{hyland91}, and with the very low amount of nebular reddening which has been inferred ($A_V \sim 0.025 - 0.05$, \citealt{mcgregor88b}).

The agreement between the CMFGEN models and the observed SED of AG~Car from the far-ultraviolet to the $L$-band is superb. The free-free emission of the wind predicted by the CMFGEN models agrees very well with the observed SED in the near-IR, suggesting that the amount of warm and hot dust ($\sim800$--1500 K) present in the close vicinity of AG~Car (1--2 \arcsec) is negligible.

\begin{table*}
\caption{Stellar and wind parameters of AG~Car during minimum epochs} 
\scriptsize
\label{agc01physpar}
\centering
\begin{tabular}{l c c c c c c c c c c}
\tableline\tableline
Epoch & V\tablenotemark{a} & BC & $\log\,(\lstar/\lsun)\tablenotemark{b}$ & \rstar & \reff   & \tstar& \teff & \mdot & \vinf  & f  \\
      &                     & (mag) & (mag)& (\rsun)& (\rsun) &  (K)  &   (K) &  (\msunyr) &  ($\kms$)  & \\ 
\tableline
1985 Jul -- 1986 Jun 	& 7.98 & -2.50    & 	6.17 & 58.5 & 78.7  & 26,450 & 22,800 & $1.9 \times 10^{-5}$ & 300 & 0.10   \\
1987 Jan -- 1990 Jun 	& 8.00 & -2.52  	&	6.17 & 59.6 & 78.7  & 26,200 & 22,800 & $1.5 \times 10^{-5}$ & 300 & 0.10   \\
1990 Dec -- 1991 Jan 	& 7.71 & -2.23   &   6.17 & 67.4 & 88.5  & 24,640 & 21,500 & $1.5 \times 10^{-5}$ & 300 & 0.10   \\
2000 Jul--2001 Jun   	& 7.63 & -2.15   &	6.17 & 85.3 & 141.6 & 21,900 & 17,000 & $ 3.7 \times 10^{-5}$ & 105 & 0.15  \\
2002 Mar--2002 Jul   	& 7.60 & -1.68	&	6.00 & 95.5 & 124.2 & 18,700 & 16,400 & $4.7 \times 10^{-5}$ & 195 & 0.25   \\
2002 Nov             	& 7.20 & -1.28   &	6.00 & 120.4 &170.4 & 16,650 & 14,000 & $6.0 \times 10^{-5}$ & 200 & 0.25   \\
2003 Jan             	& 7.03 &	-1.22   &  	6.04 & 115.2 &171.3 & 17,420 & 14,300 & $6.0 \times 10^{-5}$ & 150 & 0.25   \\
\tableline
\end{tabular}
\tablenotetext{a}{Note that the $V$ magnitude of AG Car is variable within each range of dates (see Table \ref{obsirphot1} for values for a specific date).}
\tablenotetext{b}{The photometric variability inside each range of epochs, if due to changes in $\lstar$, imply variations in the value of $\log\,(\lstar/\lsun$) listed above inside each range of epochs by $\pm 0.02$ dex.}
\end{table*}

\section{Discussion} \label{discussion}

\subsection{Chemical abundances on the surface of AG~Car and on the bipolar nebula}

The chemical abundance on the surface of AG~Car, presented in Table \ref{abund_table}, clearly shows enhancement of He, N, and Na, together with a depletion of H, C, and O compared to the solar value, which is a fingerprint that CNO-processed material is present on the stellar surface and wind of AG~Car. Since AG~Car is a prototype of the LBV class, the abundances obtained in this work quantitatively confirm the prediction that LBVs have material processed by the CNO-cycle and significant He enrichment on their surfaces (see e.g. \citealt{hd94}, \citealt{crowther97}), which put tight constraints on their evolution. 

Table \ref{abundnebula} presents the abundances obtained in this work for the surface of AG~Car and the available nebular values. Their comparison might provide clues to the evolution of AG~Car during the LBV phase, since the nebula was ejected $\sim 10^4$ years ago \citep{smith97,lamers01}. Unfortunately, nebular abundances are only reliable for N and O, since the He and C lines are weak or absent in the nebular spectrum \citep{md90,pacheco92,nota92,smith97}. The He abundance was estimated by \citet{lamers01} based on several evolutionary assumptions and on the ratio N/O obtained for the nebula. 

We support the conclusion of \citet{lamers01} that the He abundance on the surface of AG~Car is higher than in the nebula\footnote{\citet{lamers01} used the He abundance of AG~Car derived by \cite{sc94}, which is very similar to the value obtained in this work.}. In addition, we found, despite the uncertainties, that the ratio N/O is significantly higher on the stellar surface (N/O=$39^{+28}_{-18}$ by number) than in the nebula (N/O=$6 \pm 2$, \citealt{smith97}). Therefore, we conclude that the surface of the star has more CNO-processed material than the nebula. This is consistent with the idea that the massive nebula is composed of less chemically-processed outer layers which were ejected from the central star. The pre-existing interstellar bubble likely contains negligible amount of material compared to the total mass of the nebula \citep{lamers01}. Whether all the nebular mass was ejected in a single giant eruption $\sim10^4$ years ago relies on the knowledge of the mass-loss history of AG~Car earlier than and during the LBV phase, which is unknown. 

The N/O ratio obtained for AG~Car is roughly in line with the predictions of the evolutionary models by \citet{meynet03}. A star with an initial mass of $85~\msun$ reaches a similar He content as AG~Car at an age of about $3\times10^6$ years, when the evolutionary models predict N/O$\simeq26$ and N/O$\simeq20$ for the non-rotating and rotating cases, respectively. Detailed comparisons between AG~Car and evolutionary tracks, and the implications of our results for the current evolutionary scenario of massive stars will be presented in Paper II.  

In the last $10^4$ years, additional CNO-processed material could have been brought to the surface of AG~Car. The star is likely a fast rotator \citep{ghd06}, suggesting that rotational mixing could efficiently occur due to meridional circulation and shear instability \citep{maeder_araa00}. In addition, approximately $0.5~\msun$ has been removed from the star due to the steady stellar wind since the nebula was ejected, assuming an average mass-loss rate of $5\times10^{-5}~\msunyr$ during the last $10^4$ years (Paper III).

\begin{table}
\caption{Comparison between surface and nebular abundances of AG~Car }
\label{abundnebula}
\centering
\begin{tabular}{l c c c}
\hline\hline
& He/H  & N/O & Reference\\
& (number) & (number)& \\
\hline
Surface & $0.43\pm 0.08$ &  $39^{+28}_{-18}$  & this work\\
Nebula & $\sim$0.3\tablenotemark{a}  & $6 \pm 2$ & \citet{smith97} \\ 
\hline
\end{tabular}
\tablenotetext{a}{Indirectly obtained by \citet{lamers01}.}
\end{table}

\subsection{Comparison with previous works}

Fundamental parameters of AG~Car during the minimum of 2000--2003 have been presented here for the first time and, therefore, only our results for the minimum of 1985--1990 can be compared with previous works. We summarize in Table \ref{compother} the results obtained by different works.   

In general, we derived a higher $\lstar$, higher $\tstar$, lower $\teff$, and lower $\mdot$ compared to previous works. These different results are directly or indirectly related to the inclusion of wind clumping, full line blanketing, detailed treatment in non-LTE of the populations and of the radiative field in the co-moving frame, and improved treatment of the subsonic part of the wind in our analysis. The lower mass-loss rates derived by us are a direct consequence of the relatively high degree of clumping found in the wind of AG~Car ($f\simeq 0.1$), and the assumption of different $L$ and $\vinf$ by previous authors also affects the value of $\mdot$. The lower $\mdot$ and higher $\lstar$ derived by our work implies a reduction in the wind performance number $\eta=\mdot \vinf c /L$. During the minimum of 1985--1990 we found a quite modest value of $\eta \simeq 0.16$. The increase in $\tstar$ in our results can be explained by the back-warming of the base of the wind due to the inclusion of line blanketing in our modeling. The significant differences in $\teff$ also indicate that obtaining $\teff$ from simple empirical relationships, such as using the V-band flux, do not provide a correct answer for hot luminous stars with dense winds, such as AG~Car. 

The He abundance obtained in this work (He/H$=0.43 \pm 0.08$) is compatible with the value of He/H $\simeq 0.41$ determined by \citet{sc94}. All the other abundances have been derived here for the first time, including the detection of CNO-processed material on the surface of AG~Car.

The wind terminal velocity derived by us ($\vinf \simeq 300~\kms$) is higher than the value of $\vinf \simeq 250~\kms$ obtained by previous works using the FWHM of optical lines \citep{leitherer94,sc94,stahl01}, because we used the saturated absorption profile of resonance lines in the UV to infer $\vinf$, which is a more accurate diagnostic. We found that the optical lines are sensitive to the velocity law, and usually provide a lower limit to $\vinf$. \citet{guo07} derived $\beta=0.5$ for AG~Car during 1990 based on the fitting of the continuum flux, but under the assumption of a series of fixed parameters (\mdot, \rstar, \teff, \lstar, \vinf) obtained from \citet{stahl01}. All these parameters, as well as wind clumping and line blanketing, affect the shape of the continuum, and the assumed values by \citet{guo07} are significantly different than the values derived in this paper. Our detailed fits to the line profiles require a much slower velocity law ($\beta=3$) for the wind of AG~Car than that determined by \citet{guo07}. 
 
\begin{table*}
\caption{Fundamental parameters of AG~Car obtained by different works} 
\label{compother}
\centering
\begin{tabular}{l c c c c c c c c r}
\tableline
\tableline
\multicolumn{10}{c}{1989 March} \\
\cline{1-7}
\tableline
Work & log \lstar/\lsun & \tstar     & \teff  & $\mdot f^{-0.5}$      & \vinf    & $f$ & He/H & $\eta$ & $M_\mathrm{V}$\\
     &                  &    (K)     &    (K) &  ($10^{-5}~\msunyr$)   & ($\kms$) &  &   (number) & & (mag) \\
\tableline
this    & 6.17          &   26,200     &  22,800 &   4.7                & 300      & 0.1 & 0.43 & 0.17 & -8.17 \\
SCP94\tablenotemark{a}   & 6.04		    &   26,000     & \ldots &   5.6                & 250      & 1.0 & 0.41 & 0.60 & -7.7 \\
\tableline
\\
\tableline\tableline
\multicolumn{10}{c}{1990 December} \\
\cline{1-7}
\tableline
Work & log \lstar/\lsun & \tstar     & \teff  & $\mdot f^{-0.5}$      & \vinf    & $f$ & He/H & $\eta$ & $M_\mathrm{V}$\\
     &                  &    (K)     &    (K) &  ($10^{-5}~\msunyr$)   & ($\kms$) &  & (number) & &(mag) \\
\tableline
this    & 6.17          &   24,640       &  21,500 &  4.7                 & 300     & 0.1 & 0.43  & 0.17 & -8.46 \\
LEI94\tablenotemark{b}   & 6.00          &   \ldots    &  21,000 & 10.0                 & 250     & 1.0 & 1.00   & 1.42 & -8.0\\
STA01\tablenotemark{c}   & 6.00          &   \ldots    &  24,060 &  3.3                 & 225     & 1.0 & 0.41  & 0.35  & -8.0\\

\tableline
\end{tabular}
\tablenotetext{a}{\citet{sc94}.}
\tablenotetext{b}{\citet{leitherer94}.}
\tablenotetext{c}{\citet{stahl01}.}
\end{table*}

\subsection{On the nature of the changes in the bolometric luminosity during the S-Dor cycle}

The evolution of LBVs during the S-Dor cycles is widely {\it assumed} to occur roughly at constant bolometric luminosity \citep{vg79,stahl83,humphreys89,leitherer89,leitherer94,sc94,shore96,dekoter96,stahl01,vink02,walborn08}. However, the few previous works that investigated changes in the bolometric luminosity of LBVs during the S-Dor cycle (e.g. \citealt{humphreys89} for AG Car, and \citealt{walborn08} for R127) employed standard bolometric corrections of OB stars, which we argue are not realistic for LBVs, and photometry restricted to the $V$-band. For instance, the bolometric corrections obtained in this work for AG~Car are more negative by 0.15 to 0.60 mag, depending on the epoch, than those of normal B supergiants (with a similar low $\logg$) obtained using line-blanketed models such as CMFGEN \citep{crowther06} or TLUSTY \citep{lanz07}. When compared to the BCs of B supergiants provided by works which employed non-blanketed models, such as \citet{humphreys84}, the BCs derived for AG~Car are even more discrepant, being more negative by 0.40 to 0.90 mag, depending on the epoch.

Through a detailed multi-wavelength spectroscopic modeling using CMFGEN, we found in Sect. \ref{agcdistred} that the bolometric luminosity of AG~Car decreases by a factor of 1.5 ($\sim 0.2~\mathrm{dex}$) from minimum towards the maximum phase of the S-Dor cycle Sect. \ref{agcdistred}. Such a finding deserves further discussion, in particular because the variable $\lstar$ could provide a crucial constraint on the mechanism driving the S-Dor type variability. 

A reduction of the bolometric luminosity during the S-Dor cycle was also obtained by \citet{lamers95b} for the LBV S Doradus, and the result was interpreted in terms of the energy used to expand the outer layers of the star from minimum to maximum. \citet{lamers95b} found that about $10^{-3}$ to $10^{-2}~\mstar$ are involved in the expansion of S Doradus, corresponding to $\sim0.17~\msun$ (assuming $\mstar\simeq 45~\msun$, \citealt{lamers95b}).

We propose that a similar behavior is present in AG~Car. In order to obtain an order-of-magnitude estimate of the amount of mass taking part in the expansion of the star, we assume that $\lstar$ decreased from $\lstar=1.5\times10^{6}~\lsun$ to $\lstar=1.0\times10^{6}~\lsun$ on 2001 June 15 (the last epoch for which we derived $\lstar=1.5\times10^{6}~\lsun$, see Sect. \ref{agcdistred}), and remained constant until 2003 January 11. The elapsed time $\Delta t$ between 2001 June 15 and 2003 January 11 is 575 days and, therefore, the missing radiative energy during these epochs is $\Delta E_\mathrm{rad}=\Delta L \cdot \Delta t \simeq 1 \times 10^{47}~\mathrm{erg}$. Assuming that the layer is expanding from $85~\rsun$ to $115~\rsun$, and following \citet{lamers95b}, 
\begin{equation} \label{mexp}
\Delta E_\mathrm{rad}= \frac{G M_\mathrm{exp} M_\mathrm{eff}}{\rsun}\left\{\frac{1}{85}-\frac{1}{115}\right\},
\end{equation}

where $M_\mathrm{eff}=\mstar(1-\Gamma)$. Since our CMFGEN models predict that $\Gamma\simeq 0.8$, we obtain $M_\mathrm{eff} \simeq 14~\msun$ assuming $\mstar\simeq 70~\msun$ (Paper II). Using Eq. \ref{mexp}, we derive that roughly $8.3 \times 10^{-3}~\mstar \sim 0.6~\msun$ is involved in the expansion of the star during the S Dor cycle.

This result should be viewed with caution, since an uncertainty of at least a factor of 3 is present due to the various assumptions made above. The amount of $\sim0.6~\msun$ can be easily increased by a factor of 2 if we include the thermal energy released by the expanding layer (see discussion in \citealt{lamers95b}). If we also take into account that we did not analyze in this paper the epochs corresponding to the maximum in the lightcurve of AG~Car, which was only reached in 2004--2005 (Paper III), an additional amount of energy would be required, increasing by another factor of $\sim2$ the amount of mass involved in the expansion. Therefore, although detailed modeling is obviously needed, it might be possible that 0.6 to $2~\msun$ are taking part in the expansion of AG~Car during the S-Dor cycle. 

Interestingly, such an amount of mass is an order of magnitude lower than the nebular mass found around AG~Car ($\sim 15-30~\msun$, \citealt{voors00}) and that of the Homunculus nebula around Eta Car ($\sim 12-20~\msun$, \citealt{smith03b}). On the other hand, the amount of mass involved in the S-Dor type variability of AG~Car (this work) and S Doradus \citep{lamers95b} is comparable to the nebular mass found around many low-luminosity LBVs and to that of the Little Homunculus around Eta Car (see \citealt{so06} and references therein for a compilation of the amount of nebular mass found around LBVs). Such similarity might suggest a link between some of the giant outbursts of LBVs and their S-Dor cycles, and we speculate that the S-Dor cycles could actually be failed Giant Eruptions. Instead of losing several solar masses like Eta Car did in the 1840's, we suggest that during the S-Dor type variability such an amount of mass is never ejected from the star. 

\section{Conclusions } \label{conclusions}

The detailed spectroscopic analysis of AG~Car using the radiative transfer code CMFGEN has provided additional insights on the LBV phenomenon and on the S-Dor type variability. Below we summarize the main conclusions of this paper.
\begin{enumerate}
\item Following the determination of a high content of He, N, and Na, and the depletion of H, C,
and O, compared to the solar values, we infer that CNO-processed material is present on the surface of AG~Car. The actual surface is probably the peeled-off inner layers of the star, which are exposed due to the
action of a continuous stellar wind and the ejection of the outer layers due to a giant eruption that occurred 10$^4$ years ago. Since AG~Car 
is likely a fast rotator \citep{ghd06}, rotational mixing might also have played a role to increase the content of CNO-processed material on the surface.
The abundances obtained for Si, Al, Mg, Fe, Ni, Co, Cr, and Mn are consistent with a solar abundance.
\item The minimum phases of the S Dor cycle of AG~Car are not equal to each other. The consecutive minimum phases of 1985--1990
and 2000--2003 were different in duration, visual magnitude, stellar temperature, mass-loss rate, and wind terminal
velocity. We suggest that these differences arise due to different underlying stellar parameters, which cause different mass-loss rates and wind terminal velocities. In Paper II we will analyze whether the bi-stability mechanism \citep{pauldrach90,vink02} can explain this behavior.
\item The duration of the last two minimum phases of the S Dor cycle of AG~Car and their variability are related to the maximum
temperature achieved. The hotter minimum of 1985--1990 was roughly twice as long as the cooler minimum of 2000--2003. On the other hand, AG~Car was markedly more variable during 2000--2003 than during 1985--1990.
\item The comparison between the observed flux of AG~Car from the ultraviolet to the near-infrared with the model flux yielded a color excess of
$E(B-V)=0.65 \pm 0.01$, with $R_V=3.5 \pm 0.1$. The value of $R_V$ is consistent with the presence of large grains in the circumstellar
environment of AG~Car \citep{hyland91}, but it is lower than the value of $R_V=3.9$ proposed by \citet{sl94}.
\item The maximum value of the bolometric luminosity obtained during minimum phases was $\lstar=1.5\times10^{6}$L$_\odot$ for an assumed distance of
$d=6$ kpc. Excluding the uncertainty in the distance, the error in $\lstar$ due to the spectroscopic analysis is around 10\%.
\item Contrary to the current paradigm, we found that $\lstar$ decreases by a factor of 1.5 from minimum as the star moves towards visual maximum of the S-Dor cycle. Since further reduction in $\teff$ occurs during maximum (Paper III), it might be possible that $\lstar$ decreases even more during the maximum of the lightcurve. 
\item Assuming that the decrease in the bolometric luminosity of AG~Car is due the energy used to expand the outer
layers of the star \citep{lamers95b}, we found that roughly $0.6-2~\msun$ are involved in the expansion. Although that is much lower than the nebular mass found around AG~Car and that the mass of the Homunculus around Eta Car, the amount of mass involved in the S-Dor type variability of AG~Car is comparable to the nebular mass found around many low-luminosity LBVs and to that of the Little Homunculus around Eta Car. Such similarity might suggest a link between some of the giant outbursts of LBVs and their strong variability during the S-Dor cycles. We speculate that the S-Dor type instability could be a failed Giant Eruption, with the several solar masses never being released from the star. 

\end{enumerate}
 
\acknowledgments

We thank an anonymous referee for providing a very detailed list of comments and suggestions which improved the quality of the original manuscript. We are grateful to Otmar Stahl, Nolan Walborn, Edward Fitzpatrick, and Paul Crowther for kindly providing published digital spectra of AG~Car. Thanks also to Ted Gull and Krister Nielsen for insightful discussions about the nature of the absorption lines of AG~Car and for communicating results in advance of publication. We thank Michelle Fekety for proofreading the paper. JHG and AD thank Brazilian Agencies FAPESP (grant 02/11446-5) and CNPq (grant 200984/2004-7). JHG also thanks the Max-Planck-Gesellschaft (MPG) for partial financial support for this work. DJH gratefully acknowledges partial support for this work from NASA-LTSA grant NAG5-8211. JHG thanks DJH and the University of Pittsburgh for hospitality and partial support for this work. Some of the data presented in this paper were obtained from the Multimission Archive at the Space Telescope Science Institute (MAST). STScI is operated by the Association of Universities for Research in Astronomy, Inc., under NASA contract NAS5-26555. Support for MAST for non-HST data is provided by the NASA Office of Space Science via grant NAG5-7584 and by other grants and contracts. We acknowledge with thanks the variable star observations from the AAVSO International Database contributed by observers worldwide and used in this research. This research made use of the Smithsonian NASA/ADS and SIMBAD (CDS/Strasbourg) databases.

{\it Facilities:} \facility{OPD/LNA}, \facility{IUE}, \facility{ESO}, \facility{FUSE}, \facility{CTIO}.


\begin{thebibliography}{}
\bibitem[{{Bagnulo} {et~al.}(2003){Bagnulo}, {Jehin}, {Ledoux}, {Cabanac},
  {Melo}, {Gilmozzi}, \& {The ESO Paranal Science Operations Team}}]{bagnulo03}
{Bagnulo}, S., {Jehin}, E., {Ledoux}, C., {et~al.} 2003, The Messenger, 114, 10

\bibitem[{{Bandiera} {et~al.}(1989){Bandiera}, {Focardi}, {Altamore}, {Rossi},
  \& {Stahl}}]{bandiera89}
{Bandiera}, R., {Focardi}, P., {Altamore}, A., {Rossi}, C., \& {Stahl}, O.
  1989, in Astrophysics and Space Science Library, Vol. 157, IAU Colloq. 113:
  Physics of Luminous Blue Variables, ed. K.~{Davidson}, A.~F.~J. {Moffat}, \&
  H.~J.~G.~L.~M. {Lamers}, 279

\bibitem[{{Blum} {et~al.}(2000){Blum}, {Conti}, \& {Damineli}}]{bcd00}
{Blum}, R.~D., {Conti}, P.~S., \& {Damineli}, A. 2000, \aj, 119, 1860

\bibitem[{{Blum} {et~al.}(2001){Blum}, {Damineli}, \& {Conti}}]{bdc01}
{Blum}, R.~D., {Damineli}, A., \& {Conti}, P.~S. 2001, \aj, 121, 3149

\bibitem[{{Bouret} {et~al.}(2005){Bouret}, {Lanz}, \& {Hillier}}]{bouret05}
{Bouret}, J.-C., {Lanz}, T., \& {Hillier}, D.~J. 2005, \aap, 438, 301

\bibitem[{{Bouret} {et~al.}(2003){Bouret}, {Lanz}, {Hillier}, {Heap}, {Hubeny},
  {Lennon}, {Smith}, \& {Evans}}]{bouret03}
{Bouret}, J.-C., {Lanz}, T., {Hillier}, D.~J., {et~al.} 2003, \apj, 595, 1182

\bibitem[{{Brand} \& {Blitz}(1993)}]{bb93}
{Brand}, J. \& {Blitz}, L. 1993, \aap, 275, 67

\bibitem[{{Bresolin} {et~al.}(2002){Bresolin}, {Kudritzki}, {Najarro},
  {Gieren}, \& {Pietrzy{\'n}ski}}]{bresolin02}
{Bresolin}, F., {Kudritzki}, R.-P., {Najarro}, F., {Gieren}, W., \&
  {Pietrzy{\'n}ski}, G. 2002, \apjl, 577, L107

\bibitem[{{Busche} \& {Hillier}(2005)}]{bh05}
{Busche}, J.~R. \& {Hillier}, D.~J. 2005, \aj, 129, 454

\bibitem[{{Caputo} \& {Viotti}(1970)}]{caputo70}
{Caputo}, F. \& {Viotti}, R. 1970, \aap, 7, 266

\bibitem[{{Carter}(1990)}]{carter90}
{Carter}, B.~S. 1990, \mnras, 242, 1

\bibitem[{{Catala} {et~al.}(1984){Catala}, {Kunasz}, \& {Praderie}}]{catala84}
{Catala}, C., {Kunasz}, P.~B., \& {Praderie}, F. 1984, \aap, 134, 402

\bibitem[{{Clark} {et~al.}(2005){Clark}, {Larionov}, \& {Arkharov}}]{clark05}
{Clark}, J.~S., {Larionov}, V.~M., \& {Arkharov}, A. 2005, \aap, 435, 239

\bibitem[{{Conti}(1984)}]{conti84}
{Conti}, P.~S. 1984, in IAU Symp. 105: Observational Tests of the Stellar
  Evolution Theory, ed. A.~{Maeder} \& A.~{Renzini}, 233

\bibitem[{{Crowther}(1997)}]{crowther97}
{Crowther}, P.~A. 1997, in ASP Conf. Ser. 120: Luminous Blue Variables: Massive
  Stars in Transition, ed. A.~{Nota} \& H.~{Lamers}, 51

\bibitem[{{Crowther}(2006)}]{crowther06b}
{Crowther}, P.~A. 2006, in Astronomical Society of the Pacific Conference
  Series, Vol. 348, Astrophysics in the Far Ultraviolet: Five Years of
  Discovery with FUSE, ed. G.~{Sonneborn}, H.~W. {Moos}, \& B.-G. {Andersson},
  107

\bibitem[{{Crowther}(2007)}]{crowther07}
{Crowther}, P.~A. 2007, \araa, 45, 177

\bibitem[{{Crowther} {et~al.}(2006){Crowther}, {Lennon}, \&
  {Walborn}}]{crowther06}
{Crowther}, P.~A., {Lennon}, D.~J., \& {Walborn}, N.~R. 2006, \aap, 446, 279

\bibitem[{{Crowther} {et~al.}(2002){Crowther}, {Hillier}, {Evans}, {Fullerton},
  {De Marco}, \& {Willis}}]{crowther02}
{Crowther}, P.~A., {Hillier}, D.~J., {Evans}, C.~J., {et~al.} 2002, \apj, 579,
  774

\bibitem[{{Davies} {et~al.}(2005){Davies}, {Oudmaijer}, \& {Vink}}]{davies05}
{Davies}, B., {Oudmaijer}, R.~D., \& {Vink}, J.~S. 2005, \aap, 439, 1107

\bibitem[{{Davies} {et~al.}(2006){Davies}, {Oudmaijer}, \& {Vink}}]{davies06a}
{Davies}, B., {Oudmaijer}, R.~D., \& {Vink}, J.~S. 2006, in ASP Conf. Ser. 355:
  Stars with the B[e] Phenomenon, ed. M.~{Kraus} \& A.~S. {Miroshnichenko}, 173

\bibitem[{{Davies} {et~al.}(2008){Davies}, {Vink}, \& {Oudmaijer}}]{davies08}
{Davies}, B., {Vink}, J., \& {Oudmaijer}, R. 2008, in ASP Conf. Series, Vol.
  388, Mass Loss From Stars and The Evolution of Stellar Clusters, ed. A.~{de
  Koter}, L.~J. {Smith}, \& L.~B.~F.~M. {Lamers}, 71

\bibitem[{{Davies} {et~al.}(2007){Davies}, {Vink}, \& {Oudmaijer}}]{davies07}
{Davies}, B., {Vink}, J.~S., \& {Oudmaijer}, R.~D. 2007, \aap, 469, 1045

\bibitem[{{de Freitas Pacheco} {et~al.}(1992){de Freitas Pacheco}, {Damineli
  Neto}, {Costa}, \& {Viotti}}]{pacheco92}
{de Freitas Pacheco}, J.~A., {Damineli Neto}, A., {Costa}, R.~D.~D., \&
  {Viotti}, R. 1992, \aap, 266, 360

\bibitem[{{de Koter} {et~al.}(1996){de Koter}, {Lamers}, \&
  {Schmutz}}]{dekoter96}
{de Koter}, A., {Lamers}, H.~J.~G.~L.~M., \& {Schmutz}, W. 1996, \aap, 306, 501

\bibitem[{{de Wit} {et~al.}(2005){de Wit}, {Testi}, {Palla}, \&
  {Zinnecker}}]{dewit05}
{de Wit}, W.~J., {Testi}, L., {Palla}, F., \& {Zinnecker}, H. 2005, \aap, 437,
  247

\bibitem[{{Drissen} {et~al.}(2001){Drissen}, {Crowther}, {Smith}, {Robert},
  {Roy}, \& {Hillier}}]{drissen01}
{Drissen}, L., {Crowther}, P.~A., {Smith}, L.~J., {et~al.} 2001, \apj, 546, 484

\bibitem[{{Feldmeier}(1995)}]{feldmeier95}
{Feldmeier}, A. 1995, \aap, 299, 523

\bibitem[{{Fich} {et~al.}(1989){Fich}, {Blitz}, \& {Stark}}]{fich89}
{Fich}, M., {Blitz}, L., \& {Stark}, A.~A. 1989, \apj, 342, 272

\bibitem[{{Figer} {et~al.}(1998){Figer}, {Najarro}, {Morris}, {McLean},
  {Geballe}, {Ghez}, \& {Langer}}]{figer98}
{Figer}, D.~F., {Najarro}, F., {Morris}, M., {et~al.} 1998, \apj, 506, 384

\bibitem[{{Figuer{\^e}do} {et~al.}(2002){Figuer{\^e}do}, {Blum}, {Damineli}, \&
  {Conti}}]{fbdc02}
{Figuer{\^e}do}, E., {Blum}, R.~D., {Damineli}, A., \& {Conti}, P.~S. 2002,
  \aj, 124, 2739

\bibitem[{{Figuer{\^e}do} {et~al.}(2008){Figuer{\^e}do}, {Blum}, {Damineli},
  {Conti}, \& {Barbosa}}]{fbdc08}
{Figuer{\^e}do}, E., {Blum}, R.~D., {Damineli}, A., {Conti}, P.~S., \&
  {Barbosa}, C.~L. 2008, \aj, 136, 221

\bibitem[{{Fitzpatrick}(1999)}]{fitzpatrick99}
{Fitzpatrick}, E.~L. 1999, \pasp, 111, 63

\bibitem[{{Gies}(1987)}]{gies87}
{Gies}, D.~R. 1987, \apjs, 64, 545

\bibitem[{{Gr{\"a}fener} \& {Hamann}(2008)}]{grafener08}
{Gr{\"a}fener}, G. \& {Hamann}, W.-R. 2008, \aap, 482, 945

\bibitem[{{Grevesse} {et~al.}(2007){Grevesse}, {Asplund}, \&
  {Sauval}}]{grevesse07}
{Grevesse}, N., {Asplund}, M., \& {Sauval}, A.~J. 2007, Space Science Reviews,
  130, 105
  
\bibitem[{{Groh} {et~al.}(2008){Groh}, {Damineli}, \&
  {Hillier}}]{gdh08}
{Groh}, J.~H., {Damineli}, A., \& {Hillier}, D.~J. 2008, in
  Revista Mexicana de Astronomia y Astrofisica Conference Series, Vol.~33, 132  

\bibitem[{{Groh} {et~al.}(2007){Groh}, {Damineli}, \&
  {Jablonski}}]{gdj07}
{Groh}, J.~H., {Damineli}, A., \& {Jablonski}, F. 2007, \aap,
  465, 993

\bibitem[{{Groh} {et~al.}(2006){Groh}, {Hillier}, \& {Damineli}}]{ghd06}
{Groh}, J.~H., {Hillier}, D.~J., \& {Damineli}, A. 2006, \apjl, 638, L33

\bibitem[{{Gull} {et~al.}(2006){Gull}, {Kober}, \& {Nielsen}}]{gull06}
{Gull}, T.~R., {Kober}, G.~V., \& {Nielsen}, K.~E. 2006, \apjs, 163, 173

\bibitem[{{Gull} {et~al.}(2005){Gull}, {Vieira}, {Bruhweiler}, {Nielsen},
  {Verner}, \& {Danks}}]{gull05}
{Gull}, T.~R., {Vieira}, G., {Bruhweiler}, F., {et~al.} 2005, \apj, 620, 442

\bibitem[{{Guo} \& {Li}(2007)}]{guo07}
{Guo}, J.~H. \& {Li}, Y. 2007, \apj, 659, 1563

\bibitem[{{Hamann} {et~al.}(2006){Hamann}, {Gr{\"a}fener}, \&
  {Liermann}}]{hamann06}
{Hamann}, W.-R., {Gr{\"a}fener}, G., \& {Liermann}, A. 2006, \aap, 457, 1015

\bibitem[{{Herald} {et~al.}(2001){Herald}, {Hillier}, \&
  {Schulte-Ladbeck}}]{herald01}
{Herald}, J.~E., {Hillier}, D.~J., \& {Schulte-Ladbeck}, R.~E. 2001, \apj, 548,
  932

\bibitem[{{Hillier}(1987)}]{hillier87}
{Hillier}, D.~J. 1987, \apjs, 63, 947

\bibitem[{{Hillier}(1989)}]{hillier89}
{Hillier}, D.~J. 1989, \apj, 347, 392

\bibitem[{{Hillier}(1990)}]{hillier90}
{Hillier}, D.~J. 1990, \aap, 231, 116

\bibitem[{{Hillier} {et~al.}(1998){Hillier}, {Crowther}, {Najarro}, \&
  {Fullerton}}]{hillier98}
{Hillier}, D.~J., {Crowther}, P.~A., {Najarro}, F., \& {Fullerton}, A.~W. 1998,
  \aap, 340, 483

\bibitem[{{Hillier} {et~al.}(2001){Hillier}, {Davidson}, {Ishibashi}, \&
  {Gull}}]{hillier01}
{Hillier}, D.~J., {Davidson}, K., {Ishibashi}, K., \& {Gull}, T. 2001, \apj,
  553, 837

\bibitem[{{Hillier} {et~al.}(2006){Hillier}, {Gull}, {Nielsen}, {Sonneborn},
  {Iping}, {Smith}, {Corcoran}, {Damineli}, {Hamann}, {Martin}, \&
  {Weis}}]{hillier06}
{Hillier}, D.~J., {Gull}, T., {Nielsen}, K., {et~al.} 2006, \apj, 642, 1098

\bibitem[{{Hillier} {et~al.}(2003){Hillier}, {Lanz}, {Heap}, {Hubeny}, {Smith},
  {Evans}, {Lennon}, \& {Bouret}}]{hillier03}
{Hillier}, D.~J., {Lanz}, T., {Heap}, S.~R., {et~al.} 2003, \apj, 588, 1039

\bibitem[{{Hillier} \& {Miller}(1998)}]{hm98}
{Hillier}, D.~J. \& {Miller}, D.~L. 1998, \apj, 496, 407

\bibitem[{{Hillier} \& {Miller}(1999)}]{hm99}
{Hillier}, D.~J. \& {Miller}, D.~L. 1999, \apj, 519, 354

\bibitem[{{Hoekzema} {et~al.}(1992){Hoekzema}, {Lamers}, \& {van
  Genderen}}]{hoekzema92}
{Hoekzema}, N.~M., {Lamers}, H.~J.~G.~L.~M., \& {van Genderen}, A.~M. 1992,
  \aap, 257, 118

\bibitem[{{Hubeny} {et~al.}(1985){Hubeny}, {Stefl}, \& {Harmanec}}]{hubeny85}
{Hubeny}, I., {Stefl}, S., \& {Harmanec}, P. 1985, Bulletin of the Astronomical
  Institutes of Czechoslovakia, 36, 214

\bibitem[{{Humphreys}(1970)}]{humphreys70}
{Humphreys}, R.~M. 1970, \pasp, 82, 1161

\bibitem[{{Humphreys} \& {Davidson}(1994)}]{hd94}
{Humphreys}, R.~M. \& {Davidson}, K. 1994, \pasp, 106, 1025

\bibitem[{{Humphreys} {et~al.}(1989){Humphreys}, {Lamers}, {Hoekzema}, \&
  {Cassatella}}]{humphreys89}
{Humphreys}, R.~M., {Lamers}, H.~J.~G.~L.~M., {Hoekzema}, N., \& {Cassatella},
  A. 1989, \aap, 218, L17

\bibitem[{{Humphreys} \& {McElroy}(1984)}]{humphreys84}
{Humphreys}, R.~M. \& {McElroy}, D.~B. 1984, \apj, 284, 565

\bibitem[{{Hyland} \& {Robinson}(1991)}]{hyland91}
{Hyland}, A.~R. \& {Robinson}, G. 1991, Proceedings of the Astronomical Society
  of Australia, 9, 124

\bibitem[{{Johansson} {et~al.}(2005){Johansson}, {Gull}, {Hartman}, \&
  {Letokhov}}]{johansson05}
{Johansson}, S., {Gull}, T.~R., {Hartman}, H., \& {Letokhov}, V.~S. 2005, \aap,
  435, 183

\bibitem[{{Kaufer} {et~al.}(1999){Kaufer}, {Stahl}, {Tubbesing}, {Norregaard},
  {Avila}, {Francois}, {Pasquini}, \& {Pizzella}}]{kaufer99}
{Kaufer}, A., {Stahl}, O., {Tubbesing}, S., {et~al.} 1999, The Messenger, 95, 8

\bibitem[{{Lamers}(1995)}]{lamers95b}
{Lamers}, H.~J.~G.~L.~M. 1995, in Astronomical Society of the Pacific
  Conference Series, Vol.~83, IAU Colloq. 155: Astrophysical Applications of
  Stellar Pulsation, ed. J.~{Matthews}, 176

\bibitem[{{Lamers} {et~al.}(2001){Lamers}, {Nota}, {Panagia}, {Smith}, \&
  {Langer}}]{lamers01}
{Lamers}, H.~J.~G.~L.~M., {Nota}, A., {Panagia}, N., {Smith}, L.~J., \&
  {Langer}, N. 2001, \apj, 551, 764
  
\bibitem[{{Lanz} \& {Hubeny}(2007)}]{lanz07}
{Lanz}, T. \& {Hubeny}, I. 2007, \apjs, 169, 83

\bibitem[{{Leitherer} {et~al.}(1994){Leitherer}, {Allen}, {Altner}, {Damineli},
  {Drissen}, {Idiart}, {Lupie}, {Nota}, {Robert}, {Schmutz}, \&
  {Shore}}]{leitherer94}
{Leitherer}, C., {Allen}, R., {Altner}, B., {et~al.} 1994, \apj, 428, 292

\bibitem[{{Leitherer} {et~al.}(1989){Leitherer}, {Schmutz}, {Abbott}, {Hamann},
  \& {Wessolowski}}]{leitherer89}
{Leitherer}, C., {Schmutz}, W., {Abbott}, D.~C., {Hamann}, W.-R., \&
  {Wessolowski}, U. 1989, \apj, 346, 919

\bibitem[{{Maeder} \& {Meynet}(2000)}]{maeder_araa00}
{Maeder}, A. \& {Meynet}, G. 2000, \araa, 38, 143

\bibitem[{{Manfroid} {et~al.}(1991){Manfroid}, {Sterken}, {Bruch}, {Burger},
  {de Groot}, {Duerbeck}, {Duemmler}, {Figer}, {Hageman}, {Hensberge},
  {Jorissen}, {Madejsky}, {Mandel}, {Ott}, {Reitermann}, {Schulte-Ladbeck},
  {Stahl}, {Steenman}, {Vander Linden}, \& {Zickgraf}}]{manfroid91}
{Manfroid}, J., {Sterken}, C., {Bruch}, A., {et~al.} 1991, \aaps, 87, 481

\bibitem[{{Manfroid} {et~al.}(1995){Manfroid}, {Sterken}, {Cunow}, {de Groot},
  {Jorissen}, {Kneer}, {Krenzin}, {Kruijswijk}, {Naumann}, {Niehues},
  {Schoeneich}, {Sevenster}, {Vos}, \& {Vogt}}]{manfroid95}
{Manfroid}, J., {Sterken}, C., {Cunow}, B., {et~al.} 1995, \aaps, 109, 329

\bibitem[{{Marcolino} {et~al.}(2007){Marcolino}, {de Ara{\'u}jo},
  {Lorenz-Martins}, \& {Fernandes}}]{marcolino07}
{Marcolino}, W.~L.~F., {de Ara{\'u}jo}, F.~X., {Lorenz-Martins}, S., \&
  {Fernandes}, M.~B. 2007, \aj, 133, 489

\bibitem[{{Martins} {et~al.}(2005){Martins}, {Schaerer}, \&
  {Hillier}}]{martins05}
{Martins}, F., {Schaerer}, D., \& {Hillier}, D.~J. 2005, \aap, 436, 1049

\bibitem[{{McGregor} {et~al.}(1988{\natexlab{a}}){McGregor}, {Finlayson},
  {Hyland}, {Joy}, {Harvey}, \& {Lester}}]{mcgregor88b}
{McGregor}, P.~J., {Finlayson}, K., {Hyland}, A.~R., {et~al.}
  1988{\natexlab{a}}, \apj, 329, 874

\bibitem[{{McGregor} {et~al.}(1988{\natexlab{b}}){McGregor}, {Hyland}, \&
  {Hillier}}]{mcgregor88}
{McGregor}, P.~J., {Hyland}, A.~R., \& {Hillier}, D.~J. 1988{\natexlab{b}},
  \apj, 324, 1071

\bibitem[{{Meynet} \& {Maeder}(2000)}]{meynet00}
{Meynet}, G. \& {Maeder}, A. 2000, \aap, 361, 101

\bibitem[{{Meynet} \& {Maeder}(2003)}]{meynet03}
{Meynet}, G. \& {Maeder}, A. 2003, \aap, 404, 975

\bibitem[{{Mitra} \& {Dufour}(1990)}]{md90}
{Mitra}, P.~M. \& {Dufour}, R.~J. 1990, \mnras, 242, 98

\bibitem[{{Najarro}(2001)}]{najarro01}
{Najarro}, F. 2001, in Astronomical Society of the Pacific Conference Series,
  Vol. 233, P Cygni 2000: 400 Years of Progress, ed. M.~{de Groot} \&
  C.~{Sterken}, 133
  
\bibitem[{{Najarro} {et~al.}(2009){Najarro}, {Figer}, {Hillier}, {Geballe}, \&
  {Kudritzki}}]{najarro09}
{Najarro}, F., {Figer}, D.~F., {Hillier}, D.~J., {Geballe}, T.~R., \&
  {Kudritzki}, R.~P. 2009, \apj, 691, 1816

\bibitem[{{Najarro} {et~al.}(1994){Najarro}, {Hillier}, {Kudritzki}, {Krabbe},
  {Genzel}, {Lutz}, {Drapatz}, \& {Geballe}}]{najarro94}
{Najarro}, F., {Hillier}, D.~J., {Kudritzki}, R.~P., {et~al.} 1994, \aap, 285,
  573

\bibitem[{{Najarro} {et~al.}(1997){Najarro}, {Hillier}, \& {Stahl}}]{najarro97}
{Najarro}, F., {Hillier}, D.~J., \& {Stahl}, O. 1997, \aap, 326, 1117

\bibitem[{{Nielsen} {et~al.}(2005){Nielsen}, {Gull}, \& {Vieira
  Kober}}]{nielsen05}
{Nielsen}, K.~E., {Gull}, T.~R., \& {Vieira Kober}, G. 2005, \apjs, 157, 138

\bibitem[{{Nota} {et~al.}(1992){Nota}, {Leitherer}, {Clampin}, {Greenfield}, \&
  {Golimowski}}]{nota92}
{Nota}, A., {Leitherer}, C., {Clampin}, M., {Greenfield}, P., \& {Golimowski},
  D.~A. 1992, \apj, 398, 621

\bibitem[{{Nota} {et~al.}(1995){Nota}, {Livio}, {Clampin}, \&
  {Schulte-Ladbeck}}]{nota95}
{Nota}, A., {Livio}, M., {Clampin}, M., \& {Schulte-Ladbeck}, R. 1995, \apj,
  448, 788

\bibitem[{{Owocki} {et~al.}(1988){Owocki}, {Castor}, \& {Rybicki}}]{owocki88}
{Owocki}, S.~P., {Castor}, J.~I., \& {Rybicki}, G.~B. 1988, \apj, 335, 914

\bibitem[{{Pauldrach} \& {Puls}(1990)}]{pauldrach90}
{Pauldrach}, A.~W.~A. \& {Puls}, J. 1990, \aap, 237, 409

\bibitem[{{Pellerin} {et~al.}(2002){Pellerin}, {Fullerton}, {Robert}, {Howk},
  {Hutchings}, {Walborn}, {Bianchi}, {Crowther}, \& {Sonneborn}}]{pellerin02}
{Pellerin}, A., {Fullerton}, A.~W., {Robert}, C., {et~al.} 2002, \apjs, 143,
  159

\bibitem[{{Pojmanski}(2002)}]{poj02}
{Pojmanski}, G. 2002, Acta Astronomica, 52, 397

\bibitem[{{Prantzos} {et~al.}(1986){Prantzos}, {Doom}, {de Loore}, \&
  {Arnould}}]{prantzos86}
{Prantzos}, N., {Doom}, C., {de Loore}, C., \& {Arnould}, M. 1986, \apj, 304,
  695

\bibitem[{{Prinja} {et~al.}(1990){Prinja}, {Barlow}, \& {Howarth}}]{prinja90}
{Prinja}, R.~K., {Barlow}, M.~J., \& {Howarth}, I.~D. 1990, \apj, 361, 607

\bibitem[{{Puls} {et~al.}(2006){Puls}, {Markova}, {Scuderi}, {Stanghellini},
  {Taranova}, {Burnley}, \& {Howarth}}]{puls06}
{Puls}, J., {Markova}, N., {Scuderi}, S., {et~al.} 2006, \aap, 454, 625

\bibitem[{{Schmutz} {et~al.}(1989){Schmutz}, {Hamann}, \&
  {Wessolowski}}]{schmutz89}
{Schmutz}, W., {Hamann}, W.-R., \& {Wessolowski}, U. 1989, \aap, 210, 236

\bibitem[{{Schulte-Ladbeck} {et~al.}(1994){Schulte-Ladbeck}, {Clayton},
  {Hillier}, {Harries}, \& {Howarth}}]{sl94}
{Schulte-Ladbeck}, R.~E., {Clayton}, G.~C., {Hillier}, D.~J., {Harries}, T.~J.,
  \& {Howarth}, I.~D. 1994, \apj, 429, 846

\bibitem[{{Schulte-Ladbeck} {et~al.}(1997){Schulte-Ladbeck}, {Schmid}, {Meade},
  {Harries}, {Lupie}, \& {Bjorkman}}]{sl97}
{Schulte-Ladbeck}, R.~E., {Schmid}, H.~M., {Meade}, M.~R., {et~al.} 1997, in
  ASP Conf. Ser. 120: Luminous Blue Variables: Massive Stars in Transition, ed.
  A.~{Nota} \& H.~{Lamers}, 113

\bibitem[{{Shore} {et~al.}(1996){Shore}, {Altner}, \& {Waxin}}]{shore96}
{Shore}, S.~N., {Altner}, B., \& {Waxin}, I. 1996, \aj, 112, 2744

\bibitem[{{Smith} {et~al.}(1994){Smith}, {Crowther}, \& {Prinja}}]{sc94}
{Smith}, L.~J., {Crowther}, P.~A., \& {Prinja}, R.~K. 1994, \aap, 281, 833

\bibitem[{{Smith} {et~al.}(1997){Smith}, {Stroud}, {Esteban}, \&
  {Vilchez}}]{smith97}
{Smith}, L.~J., {Stroud}, M.~P., {Esteban}, C., \& {Vilchez}, J.~M. 1997,
  \mnras, 290, 265

\bibitem[{{Smith} {et~al.}(2003){Smith}, {Gehrz}, {Hinz}, {Hoffmann}, {Hora},
  {Mamajek}, \& {Meyer}}]{smith03b}
{Smith}, N., {Gehrz}, R.~D., {Hinz}, P.~M., {et~al.} 2003, \aj, 125, 1458

\bibitem[{{Smith} \& {Owocki}(2006)}]{so06}
{Smith}, N. \& {Owocki}, S.~P. 2006, \apjl, 645, L45

\bibitem[{{Spoon} {et~al.}(1994){Spoon}, {de Koter}, {Sterken}, {Lamers}, \&
  {Stahl}}]{spoon94}
{Spoon}, H.~W.~W., {de Koter}, A., {Sterken}, C., {Lamers}, H.~J.~G.~L.~M., \&
  {Stahl}, O. 1994, \aaps, 106, 141

\bibitem[{{Stahl}(1986)}]{stahl86}
{Stahl}, O. 1986, \aap, 164, 321

\bibitem[{{Stahl} {et~al.}(2001){Stahl}, {Jankovics}, {Kov{\' a}cs}, {Wolf},
  {Schmutz}, {Kaufer}, {Rivinius}, \& {Szeifert}}]{stahl01}
{Stahl}, O., {Jankovics}, I., {Kov{\' a}cs}, J., {et~al.} 2001, \aap, 375, 54

\bibitem[{{Stahl} {et~al.}(1999){Stahl}, {Kaufer}, \& {Tubbesing}}]{stahl99}
{Stahl}, O., {Kaufer}, A., \& {Tubbesing}, S. 1999, in ASP Conf. Ser. 188:
  Optical and Infrared Spectroscopy of Circumstellar Matter, 331
  
\bibitem[{{Stahl} {et~al.}(1993){Stahl}, {Mandel}, {Wolf}, {Gaeng}, {Kaufer},
  {Kneer}, {Szeifert}, \& {Zhao}}]{stahl93}
{Stahl}, O., {Mandel}, H., {Wolf}, B., {et~al.} 1993, \aaps, 99, 167

\bibitem[{{Stahl} {et~al.}(1983){Stahl}, {Wolf}, {Klare}, {Cassatella},
  {Krautter}, {Persi}, \& {Ferrari-Toniolo}}]{stahl83}
{Stahl}, O., {Wolf}, B., {Klare}, G., {et~al.} 1983, \aap, 127, 49

\bibitem[{{Sterken}(1983)}]{sterken83}
{Sterken}, C. 1983, The Messenger, 33, 10

\bibitem[{{Sterken} {et~al.}(1996){Sterken}, {Jones}, {Vos}, {Zegelaar}, {van
  Genderen}, \& {de Groot}}]{sterken96ibvs}
{Sterken}, C., {Jones}, A., {Vos}, B., {et~al.} 1996, Informational Bulletin on
  Variable Stars, 4401, 1

\bibitem[{{Sterken} {et~al.}(1993){Sterken}, {Manfroid}, {Anton}, {Barzewski},
  {Bibo}, {Bruch}, {Burger}, {Duerbeck}, {Duemmler}, {Heck}, {Hensberge},
  {Hiesgen}, {Inklaar}, {Jorissen}, {Juettner}, {Kinkel}, {Zongli}, {Mekkaden},
  {Ng}, {Niarchos}, {Puttmann}, {Szeifert}, {Spiller}, {van Dijk}, {Vogt}, \&
  {Wanders}}]{sterken93}
{Sterken}, C., {Manfroid}, J., {Anton}, K., {et~al.} 1993, \aaps, 102, 79

\bibitem[{{Sterken} {et~al.}(1995){Sterken}, {Stahl}, {Wolf}, {Szeifert}, \&
  {Jones}}]{sterken95}
{Sterken}, C., {Stahl}, O., {Wolf}, B., {Szeifert}, T., \& {Jones}, A. 1995,
  \aap, 303, 766

\bibitem[{{van Genderen}(1979)}]{vg79}
{van Genderen}, A.~M. 1979, \aaps, 38, 381

\bibitem[{{van Genderen}(1982)}]{vg82}
{van Genderen}, A.~M. 1982, \aap, 112, 61

\bibitem[{{van Genderen}(2001)}]{vg01}
{van Genderen}, A.~M. 2001, \aap, 366, 508

\bibitem[{{van Genderen} {et~al.}(1997{\natexlab{a}}){van Genderen}, {de
  Groot}, \& {Sterken}}]{vg97}
{van Genderen}, A.~M., {de Groot}, M., \& {Sterken}, C. 1997{\natexlab{a}},
  \aaps, 124, 517

\bibitem[{{van Genderen} {et~al.}(1997{\natexlab{b}}){van Genderen}, {Sterken},
  \& {de Groot}}]{vg97b}
{van Genderen}, A.~M., {Sterken}, C., \& {de Groot}, M. 1997{\natexlab{b}},
  \aap, 318, 81

\bibitem[{{van Genderen} {et~al.}(1988){van Genderen}, {The}, {Augusteijn},
  {Engelsman}, {van der Grift}, {Prein}, {Remijn}, {Steeman}, \& {van
  Weeren}}]{vg88}
{van Genderen}, A.~M., {The}, P.~S., {Augusteijn}, T., {et~al.} 1988, \aaps,
  74, 453

\bibitem[{{van Genderen} {et~al.}(1990){van Genderen}, {The}, {Heemskerk},
  {Heynderickx}, {van Kampene}, {Kraakman}, {Larsen}, {Remijn}, {Wanders}, \&
  {van Weeren}}]{vg90}
{van Genderen}, A.~M., {The}, P.~S., {Heemskerk}, M., {et~al.} 1990, \aaps, 82,
  189

\bibitem[{{Vink} \& {de Koter}(2002)}]{vink02}
{Vink}, J.~S. \& {de Koter}, A. 2002, \aap, 393, 543

\bibitem[{{Viotti}(1971)}]{viotti71}
{Viotti}, R. 1971, \pasp, 83, 170

\bibitem[{{Viotti} {et~al.}(1991){Viotti}, {Baratta}, {Rossi}, \& {di
  Fazio}}]{viotti91}
{Viotti}, R., {Baratta}, G.~B., {Rossi}, C., \& {di Fazio}, A. 1991, in IAU
  Symposium, Vol. 143, Wolf-Rayet Stars and Interrelations with Other Massive
  Stars in Galaxies, ed. K.~A. {van der Hucht} \& B.~{Hidayat}, 499

\bibitem[{{Viotti} {et~al.}(1993){Viotti}, {Polcaro}, \& {Rossi}}]{viotti93}
{Viotti}, R., {Polcaro}, F.~V., \& {Rossi}, C. 1993, \aap, 276, 432

\bibitem[{{Voors} {et~al.}(2000){Voors}, {Waters}, {de Koter}, {Bouwman},
  {Morris}, {Barlow}, {Sylvester}, {Trams}, \& {Lamers}}]{voors00}
{Voors}, R.~H.~M., {Waters}, L.~B.~F.~M., {de Koter}, A., {et~al.} 2000, \aap,
  356, 501

\bibitem[{{Walborn} \& {Fitzpatrick}(2000)}]{wf2000}
{Walborn}, N.~R. \& {Fitzpatrick}, E.~L. 2000, \pasp, 112, 50

\bibitem[{{Walborn} {et~al.}(2002){Walborn}, {Fullerton}, {Crowther},
  {Bianchi}, {Hutchings}, {Pellerin}, {Sonneborn}, \& {Willis}}]{walborn02}
{Walborn}, N.~R., {Fullerton}, A.~W., {Crowther}, P.~A., {et~al.} 2002, \apjs,
  141, 443

\bibitem[{{Walborn} {et~al.}(2008){Walborn}, {Stahl}, {Gamen}, {Szeifert},
  {Morrell}, {Smith}, {Howarth}, {Humphreys}, {Bond}, \& {Lennon}}]{walborn08}
{Walborn}, N.~R., {Stahl}, O., {Gamen}, R.~C., {et~al.} 2008, \apjl, 683, L33

\bibitem[{{Whitelock} {et~al.}(1983){Whitelock}, {Carter}, {Roberts},
  {Whittet}, \& {Baines}}]{whitelock83}
{Whitelock}, P.~A., {Carter}, B.~S., {Roberts}, G., {Whittet}, D.~C.~B., \&
  {Baines}, D.~W.~T. 1983, \mnras, 205, 577

\bibitem[{{Wolf}(1989)}]{wolf89}
{Wolf}, B. 1989, \aap, 217, 87

\bibitem[{{Wolf} \& {Stahl}(1982)}]{wolf82}
{Wolf}, B. \& {Stahl}, O. 1982, \aap, 112, 111

\end{thebibliography}
\end{document}